\documentclass[12pt]{article}
\usepackage{amsmath}
\usepackage{multirow}
\usepackage{amsfonts}
\usepackage{amssymb,graphics,psfrag}
\usepackage{array,epsfig,multirow,graphicx}
\usepackage{comment}
\usepackage{mathrsfs, dsfont}
\usepackage{color}
\usepackage[bf]{caption}
\usepackage[pdftex,
    pdftitle={Explicit de Sitter Flux Vacua for Global String Models with Chiral Matter},
    pdfauthor={Michele Cicoli, Denis Klevers, Sven Krippendorf, Christoph Mayrhofer, Fernando Quevedo, Roberto Valandro},
    pdfsubject={},
    bookmarksopen, bookmarksnumbered, bookmarksopenlevel=2]{hyperref}
\usepackage{collref}

\def\hybrid{\topmargin -20pt    \oddsidemargin 0pt
        \headheight 0pt \headsep 0pt
        \textwidth 6.25in
        \textheight 9 in
        \marginparwidth .875in
        \parskip 5pt plus 1pt
          \jot = 1.5ex
  }
\hybrid

\numberwithin{equation}{section}

\newcommand{\be}{\begin{equation}}
\newcommand{\ee}{\end{equation}}
\newcommand{\bea}{\begin{eqnarray}}
\newcommand{\eea}{\end{eqnarray}}

\newcommand{\mc}{\mathcal}
\newcommand{\nn}{\nonumber}

\newcommand{\beq}{\begin{equation}}
\newcommand{\eeq}{\end{equation}}
\newcommand{\vo}{\mathcal{V}}

\newcommand{\cD}{\mathcal{D}}
\newcommand{\cL}{\mathcal{L}}

\newcommand{\cK}{\mathcal{K}}

\newcommand{\cF}{\mathcal{F}}

\newcommand{\Z}{\mathds{Z}}

\newcommand{\cref}{{\bf [check ref]}}

\newcommand{\norm}[1]{\lVert #1\rVert}

\def\blfootnote{\xdef\@thefnmark{}\@footnotetext}
\long\def\symbolfootnote[#1]#2{\begingroup%
\def\thefootnote{\fnsymbol{footnote}}\footnote[#1]{#2}\endgroup}

\begin{document}

\baselineskip=15pt

\begin{titlepage}
\begin{flushright}
\parbox[t]{1.08in}{UPR-1256-T}\\
\parbox[t]{1.38in}{DAMTP-2013-69}
\end{flushright}

\begin{center}

\vspace*{ 1.0cm}

{\Large \bf Explicit de Sitter Flux Vacua for Global String Models with Chiral Matter}

\vskip 1.0cm

\renewcommand{\thefootnote}{}
\begin{center}
{Michele Cicoli${}^{1,2,3},$ Denis Klevers${}^4,$ Sven Krippendorf${}^5,$ Christoph Mayrhofer${}^6,$ \\ Fernando Quevedo${}^{3,7},$ Roberto Valandro${}^{3,8}$}
\end{center}
\vskip .2cm
\renewcommand{\thefootnote}{\arabic{footnote}}

{\it \small $^1$ Dipartimento di Fisica e Astronomia, Universit\`a di Bologna, \\ via Irnerio 46, 40126 Bologna, Italy. \\
$^2$ INFN, Sezione di Bologna, Italy.\\
$^3$ ICTP, Strada Costiera 11, Trieste 34014, Italy.\\
$^4$ {Department of Physics and Astronomy,\\
University of Pennsylvania, Philadelphia, PA 19104-6396, USA.\\
$^5$ Bethe Center for Theoretical Physics and Physikalisches Institut der\\ Universit\"at Bonn, Nussallee 12, 53115 Bonn, Germany.\\
$^6$ Institut f\"ur Theoretische Physik, Universit\"at Heidelberg,\\ Philosophenweg 19,  69120 Heidelberg, Germany.\\
$^7$ DAMTP, University of Cambridge, Wilberforce Road, Cambridge, CB3 0WA, UK.\\
$^8$ INFN, Sezione di Trieste, Italy.}\\[0.5cm]}

{\small mcicoli@ictp.it, klevers@sas.upenn.edu, krippendorf@th.physik.uni-bonn.de, C.Mayrhofer@thphys.uni-heidelberg.de, F.Quevedo@damtp.cam.ac.uk, rvalandr@ictp.it}


\end{center}


\begin{center} {\bf ABSTRACT} \end{center}
We address the open question of performing an explicit stabilisation of all closed string moduli (including dilaton, complex structure and K\"ahler moduli)
in fluxed type IIB Calabi-Yau compactifications with chiral matter.
Using toric geometry we construct Calabi-Yau manifolds with del Pezzo singularities. D-branes located at such singularities can support the Standard Model gauge group and matter content. In order to control complex structure moduli stabilisation we consider Calabi-Yau manifolds which exhibit a discrete symmetry that reduces the effective number of complex structure moduli. We calculate the corresponding periods in the symplectic basis of invariant three-cycles  and find explicit flux vacua for concrete examples. We compute the values of the flux superpotential and the string coupling at these vacua. 
Starting from these explicit complex structure solutions, we obtain AdS and dS minima where the K\"ahler moduli are stabilised by a mixture of D-terms, non-perturbative and perturbative $\alpha'$ corrections as in the LARGE Volume Scenario.
In the considered example the visible sector lives at a dP$_6$ singularity which can be higgsed to the phenomenologically interesting class of models at the dP$_3$ singularity.
\end{titlepage}

\tableofcontents

\section{Introduction}

Constructing realistic string vacua is technically challenging as top-down consistency conditions of string compactifications and bottom-up experimental constraints on the effective field theory lead to strict limitations for any string model. In spite of the difficulty of the task, the bottom-up approach to string model building~\cite{hep-th/0005067} allows to address various requirements once at a time. Moreover, over the past decade various successful mechanisms, in particular in the context of type~IIB string theory, have been established which allow to address moduli stabilisation and realistic D-brane model building~\cite{Ibanez:2012zz}. The former can be achieved by flux stabilisation of the complex structure moduli and the dilaton (following for example~\cite{Giddings:2001yu}), and a mixture of perturbative and non-perturbative corrections in the effective field theory can be used to stabilise the K\"ahler moduli~\cite{Kachru:2003aw,Balasubramanian:2005zx}. As far as model building is concerned,
D-branes at singularities provide an interesting avenue towards realistic models of particle physics which
have been shown to realise extensions of the MSSM with interesting flavour structure (see for instance~\cite{1212.0555, 0810.5660,1002.1790,1102.1973,1106.6039}). Similarly local F-theory models provide another interesting class of models with realistic properties~\cite{1212.0555,Weigand:2010wm,Heckman:2010bq,Vafa:2009se,Wijnholt:2008db}.

For a fully realistic string compactification, all of these successful mechanisms have to be combined in a consistent string set-up. Recently~\cite{Cicoli:2011qg,Cicoli:2012vw,1304.0022,1304.2771} we addressed the question of combining K\"ahler moduli stabilisation with D-brane model building and we continue this program here by also implementing explicitly flux stabilisation of complex structure moduli. Some of the previous accomplishments include:
\begin{itemize}
\item Using toric geometry, we constructed explicit compact Calabi-Yau (CY) orientifold compactifications \cite{Mayrhofer2010,Cicoli:2011it}\nocite{Gao:2013pra}.
\item Chiral matter arises from fractional D3/D7-branes at del Pezzo singularities.
These singularities are mapped onto each other by the orientifold involution, realising an invariant setting.
The D-brane configuration at these singularities allows for a visible sector including the Standard Model gauge group.
\item The geometries contain an additional `small' (del Pezzo) divisor to generate a non-perturbative superpotential for moduli stabilisation.
\item An additional four-cycle class controls the overall size of the compactification which, overall,
leads to CY manifolds with $h^{1,1}\geq 4$.
\item We performed a full classification of all models of this type with $h^{1,2}\geq 5\geq h^{1,1}\geq 4$.
\item In explicit examples we showed that all known consistency conditions can be satisfied (D-brane tadpole cancellation, Freed-Witten anomalies and K-theory constraints).
\item The K\"ahler moduli can be stabilised as in the LARGE Volume Scenario (LVS), leading to explicit realisations of de Sitter minima by including non-vanishing matter F-term contributions induced by D-term fixing. Supersymmetry is softly broken and can lead to TeV-scale soft masses.
\end{itemize}
The aim of this paper is to add an \emph{additional} layer to the above: the explicit flux stabilisation of dilaton and complex structure moduli, while keeping the previously listed properties of the compactification untouched. Typically the accessible CY manifolds in this class of models have a few K\"ahler moduli but many complex structure moduli, rendering the analysis of the potential of the complex structure moduli very difficult. However, there can be additional symmetries in the complex structure moduli space such that the effective number of complex structure moduli is reduced \cite{hep-th/0312104,hep-th/0404257}. In particular we are interested in a subclass of such CY manifolds: those that allow for the Greene-Plesser (GP) construction of the mirror manifold~\cite{Greene:1990ud}.

We actively search for such symmetries in the models we previously classified as phenomenologically interesting and identify several examples that allow for this mechanism. In these examples, we explicitly calculate the leading and sub-leading contribution to the prepotential for the special K\"ahler complex structure moduli space (in the mirror space language, these are polynomial terms and instanton corrections).
In one of these examples, we then search for flux minima which set two phenomenologically relevant parameters: the vacuum expectation value (VEV) of the flux superpotential $W_0$ and the string coupling $g_s$. 
Due to computational challenges we only have so far a small number of trustable solutions. Nevertheless we find agreement with previous discussions on the approximately uniform distribution of $W_0$ and $g_s$ in the patch of moduli space with large complex structure moduli~\cite{Denef:2004ze,1208.3208,MartinezPedrera:2012rs}. We then combine these results with the explicit stabilisation of the K\"ahler moduli. Hence we work out the first explicit example that allows for closed string moduli stabilisation to de Sitter space and a visible sector on a dP$_6$ singularity. We finally show that the visible sector gauge theory can be related by appropriate VEVs to the interesting class of gauge theories arising from the dP$_3$ singularity.

The rest of this paper is organised as follows.
In Section~\ref{sec:cs} we discuss how to explicitly calculate the dilaton and complex structure contribution to the effective supergravity potential (i.e.~the prepotential) and present an example for this calculation. For this example, we present a chiral D-brane setup and check for UV consistency conditions in Section~\ref{sec:ExplicitModel}. We stabilise first the dilaton and complex structure moduli, including a short discussion on the statistics of the flux minima, and then the K\"ahler moduli in Section~\ref{sec:stabilisation} before concluding with Section~\ref{sec:Conclusions}.
In Appendix~\ref{app:Review} we present more details on the analysis of the complex structure moduli space and in Appendix~\ref{app:MoreEx} we list further examples that obey our search criteria.

\section{Effective field theory for complex structure moduli}
\label{sec:cs}

In this section we explain how one can concretely calculate the K\"ahler potential
and the flux superpotential for the complex structure moduli.

As we review in the following, the computation of both
quantities reduces to the problem of finding the periods of a family
of CY three-folds $M_3$. The periods are completely encoded in
the holomorphic prepotential $F$ on the complex structure moduli space
of $M_3$. The structure and computation of $F$ is addressed most
conveniently in the context of mirror symmetry
and $N=2$ special geometry, of which we briefly review the basic
notions in Section \ref{sec:periods} and refer to Appendix \ref{app:Review} for 
a more detailed summary. 
After general remarks,
we explicitly compute $F$ and the periods for a concrete CY three-fold
$M_3$ in Section \ref{sec:example1}. This three-fold exhibits the
necessary geometrical properties for constructing a phenomenologically
promising D-brane setup, cf.~Section \ref{sec:ExplicitModel},
for which we perform moduli stabilisation explicitly in Section
\ref{sec:stabilisation}. 

For an earlier discussion of the $N=2$ special 
geometry perspective on type II flux compactifications, that 
parallels parts of the presentation below, along with an
analysis of their flux vacua, at generic points and explicitly for certain 
fluxes at degeneration points of the Calabi-Yau geometries, see 
\cite{Curio:2000sc}. 

We consider Calabi-Yau three-folds $M_3$ with a key property, i.e.
the presence of a symmetry on its complex structure moduli
space, that allows to fix a large number of complex structure
moduli without doing any computation, as explained in Section
\ref{sec:ReducingCSM}. This
makes the explicit computation of the periods of $M_3$ (performed
in Section \ref{sec:example}) technically feasible.
Explicit computations of periods for other examples of CY three-folds
with such symmetries on their complex structure moduli spaces are
presented in Appendix \ref{app:MoreEx}.

Before delving into the details of the computations, we begin
by introducing some notation and the basic idea behind the stabilisation
of dilaton and complex structure moduli which we denote as $\tau=C_0+ig_s^{-1}$ and $u^i$, $i=1,\ldots, h^{1,2}(M_3)$.
The quantised RR- and NSNS-fluxes, $F_3=dC_2$ and $H_3=dB_2$,
conveniently combined into $G_3=F_3-\tau H_3$, induce the perturbative flux superpotential
\beq \label{eq:Wflux}
W_{\rm flux}(u^i,\tau)=\int_{M_3} \Omega\wedge G_3\,.
\eeq
It depends on the moduli
$\tau$, $u^i$, where the dependence on the $u^i$ arises from the holomorphic three-form $\Omega$ on $M_3$. The K\"ahler potential
for these moduli reads
\beq \label{eq:Kcs+tau}
K=-\ln\left(i\int\Omega\wedge \bar{\Omega}\right)-\ln{(-i (\tau-\bar{\tau}))}\,.
\eeq
Further dependence on $\tau$ and $u^i$ appears in the K\"ahler and superpotential upon the inclusion of open strings and $\alpha'$-corrections. From the field theory analysis, these effects turn out to be sub-leading in the process of moduli stabilisation, and so we shall neglect them (see for instance~\cite{Balasubramanian:2005zx}).

Solving the F-term conditions following from \eqref{eq:Wflux} and \eqref{eq:Kcs+tau}:
\beq \label{eq:F-term}
D_{u^i}W=(\partial_{u^i}+K_{u^i})W=0\,,\qquad D_{\tau}W=(\partial_{\tau}+K_{\tau})W=0
\eeq
generically fixes both $\tau$ and $u^i$. We will demonstrate this explicitly in 
Section~\ref{sec:stabilisation} using the results from Section 
\ref{sec:example}. Notice that even though these are $h^{1,2}+ 1$ first order 
equations with the same number of unknowns, there is no a priory guarantee of a 
solution since they are non-holomorphic. However there are $2h^{1,2}+ 2$ 
flux parameters which generically implies the existence of many solutions which 
need to be determined numerically. This is of course the main source for the 
existence of the landscape of type IIB string vacua.

\subsection{K\"ahler potential and flux superpotential}
\label{sec:periods}

In this subsection,
we present a brief review of the special geometry of
the complex structure moduli space of a CY three-fold and mirror
symmetry, following mainly \cite{Klemm:2005tw}.

First, we notice that the complex structure moduli
dependence in \eqref{eq:Wflux} and \eqref{eq:Kcs+tau} is captured
entirely by the holomorphic three-form $\Omega$.
In order to compute the dependence of $\Omega$ on the $u^i$
explicitly, we introduce the integral symplectic basis $(\alpha_K,\beta^K)$ of
$H^3(M_3,\mathbb{Z})$ where  $K=0,\ldots, h^{2,1}(M_3)$. Then, we can expand
the holomorphic three-form $\Omega$ as
\beq \label{eq:periodExp}
\Omega=X^K\alpha_K-\mathcal{F}_K\beta^K\,.
\eeq
The coefficients are called periods of $\Omega$. They are conveniently combined into the period vector
$\Pi^t=(X^K,\mathcal{F}_K)$.
The $X^K$'s are projective coordinates on the complex structure moduli
space which has complex dimension $h^{2,1}$.
The periods $\cF_K$'s are functions of the $X^K$'s. They can be obtained (in a proper symplectic frame) as
the partial derivatives of a single holomorphic function $\cF$, the projective
prepotential, as
\begin{equation}
	\mathcal{F}_L=\frac12\frac{\partial}{\partial X^L}(X^K\mathcal{F}_K)\equiv \frac{\partial}{\partial X^L}
	\mathcal{F}\,.
\label{eq:prepotProjective}
\end{equation}
The prepotential is a homogeneous function of degree two in the $X^K$'s.\footnote{Globally, $\mathcal{F}$ is a
section of the square of the vacuum line bundle $\mathcal{L}^2$ over
the complex structure moduli space.}
We introduce the function
\begin{equation}
	F(u^i)=(X^0)^{-2}\mathcal{F}(X^K)\, ,
\label{eq:prepotAffine}
\end{equation}
(that we still call `prepotential') where we defined the flat coordinates:\footnote{We deviate here from the usual convention in the mirror symmetry literature,
where the flat coordinates are denoted by $t^i$, in order to avoid confusions with the K\"ahler moduli of $M_3$.}
\beq \label{eq:flatcoord}
u^i=\frac{X^i}{X^0}\,,\qquad i\neq 0 \, .
\eeq
In terms of the prepotential, we can express the period vector as
\beq \label{eq:PiXF}
	\Pi=\begin{pmatrix}
	X^0\\
	X^i\\
	\mathcal{F}_0\\
	\mathcal{F}_i
	\end{pmatrix}\cong\begin{pmatrix}
						1\\
						u^i\\
						2F-u^i\partial_i F\\
						\partial_i F
				\end{pmatrix}\,.
\eeq
where we used the split $I=(0,i)$ for $i=1,\ldots, h^{1,2}(M_3)$ and in
the second equality we normalised $\Omega$ by choosing $X^0=1$.

The superpotential and the K\"ahler potential $K_{\rm{cs}}$ for the
complex structure moduli (cf.~\eqref{eq:Wflux} and ~\eqref{eq:Kcs+tau}) can now be rewritten
as
\bea \label{eq:KaehlerpotCS}
W_{\rm flux}&=&\int G_3\wedge\Omega=(f-\tau h)^t\cdot\Pi,\\
	\nonumber K_{\rm{cs}}&=&-\ln\left(i\int\Omega\wedge \bar \Omega\right)=-\ln\left(i\Pi^\dagger\cdot\Sigma\cdot\Pi\right)\\
	&=&-\ln \left(2i\,\text{Im}(\bar{X}^K\mathcal{F}_K)\right)
	=-\ln\left(i\norm{X^0}^2(2F-2\bar{F}-(u-\bar u)^i(F+\bar F)_i)\right)\,,\nn
\eea
where $``\cdot"$ is the ordinary scalar product and $\Sigma$
is the standard symplectic matrix
\beq
	\Sigma=\begin{pmatrix}
		0 & \mathbf{1}\\
		-\mathbf{1} & 0
	\end{pmatrix}\, .
\eeq
Moreover, the background fluxes can be expanded in the
symplectic basis $(\alpha_K,\beta^K)$ as
\beq
G_3=(M^K -\tau \tilde M^K)\alpha_K-(N_K - \tau \tilde N_K)\beta^K\,.
\eeq
We have introduced the vectors $f=\Sigma \cdot (M^K,N_K)^t$ and
$h=\Sigma \cdot(\tilde{M}^K,\tilde{N}_K)^t$ 
which encode the RR- and
NSNS-fluxes.
The flux contribution to the D3-tadpole reads
\be
Q_{D3}=\frac{1}{(2\pi)^4 (\alpha')^2}\int_{M_3} H_3\wedge F_3=  M^K \tilde N_K-\tilde M^K N_K=-h^t\cdot\Sigma\cdot f \,.
\label{eq:D3tad}
\ee

From \eqref{eq:PiXF} it is now manifest that the key quantity encoding all the
information of the periods $\Pi$ is the holomorphic prepotential $F$. Its `large
complex structure' expansion on $M_3$ in terms  of the coordinates $u^i$
defined in \eqref{eq:flatcoord} is given by (see for example~\cite{Klemm:2005tw})\footnote{We refer to Equations \eqref{eq:periodVectorCY3LV} and \eqref{eq:flux_ori} in Appendix~\ref{app:Review} for explicit
expressions for $\Pi$ and $W_{\text{flux}}$ in terms of the quantities appearing in \eqref{eq:pre_largeV}.}
\begin{equation} \label{eq:pre_largeV}
F = -\tfrac{1}{3!} \cK_{ijk}\, u^i u^j u^k - \tfrac{1}{2!}\cK_{ij}\, u^i u^j + \cK_{i} u^i + \tfrac12 \cK_0
 + \sum_{\beta} n_\beta\,\text{Li}_3(q^\beta)\, .
\end{equation}
This general form of \eqref{eq:pre_largeV}
is explained most easily from the point of view of the
mirror CY three-fold $\tilde{M}_3$ and the A-model perspective as the `large radius/large volume expansion'. Here $F$ is the prepotential on the quantum
corrected K\"ahler moduli space of $\tilde{M}_3$, with its K\"ahler moduli related to the flat coordinates \eqref{eq:flatcoord}
on $M_3$ via the mirror map
\beq \label{eq:t^iIIA}
u^i_{\rm B-model}\qquad \leftrightarrow\qquad t^i_{\rm A-model}=\int_{\beta^i}(J+i B_2)\,.
\eeq
The two-form $B_2$ denotes the NSNS $B$-field, $J$ is the K\"ahler form on
$\tilde{M}_3$ and the $\beta^i$ are the generators of
$H_{2}(\tilde{M}_3,\mathbb{Z})$.
The classical terms in \eqref{eq:pre_largeV} are determined by
the classical intersections on $\tilde{M}_3$ according to \cite{Mayr:2000as} (see also \cite{Grimm:2009ef}
for generalisations to higher dimensions):
\begin{align} \label{eq:classic_terms_threefold2}
  &\cK_{ijk} = \int_{\tilde M_3} J_i \wedge J_j \wedge J_k\ ,\qquad
  &&\mathcal{K}_{ij}=-\frac{1}{2}\int_{\tilde{M}_3}
	J_i\wedge J_j^2\,\,\,\,\text{mod}\, \mathbb{Z}\ , \\
  &\cK_j = \frac{1}{2^2 3!}\int_{\tilde M_3} c_2({\tilde M_3})\wedge J_j\,\,\,\,\text{mod}\, \mathbb{Z}\ , \qquad
  &&\cK_0= \frac{\zeta(3)}{(2 \pi i)^3}\int_{\tilde M_3} c_3({\tilde M_3}) \,.\nn
\end{align}
Here $J_i$ denote $(1,1)$-forms that are  dual to the generators  $K_i$
of the K\"ahler cone of $\tilde{M}_3$ and
$c_{2,3}({\tilde M_3})$ denote the second (third) Chern class of the
mirror CY.
Note that the coefficients $\mathcal{K}_{ij}$ and $\mathcal{K}_i$
of the sub-leading terms in the $u^i$ are only fixed modulo
integers. Fixing them corresponds to a particular choice of integral
symplectic basis $(\alpha_K,\beta^L)$.\footnote{The freedom in the choice of such a basis
is given by the symplectic transformations $Sp(2+2h^{1,2},\Z)$, w.r.t.~which the period
vector $\Pi$ transforms in the fundamental representation. For the prepotential
\eqref{eq:pre_largeV} with the classical terms \eqref{eq:classic_terms_threefold2} the period
vector $\Pi$ is obtained in one particular choice of basis.} We refer to Appendix
\ref{app:Review} for more details on the derivation of
\eqref{eq:classic_terms_threefold2}.
The last term in \eqref{eq:pre_largeV} are the
string worldsheet instanton corrections to the A-model. Here we employed
the notation $q^\beta = e^{2\pi id_j u^j}$ for a class $\beta=d_i\beta^i$
with degrees $d_i\in\mathbb{Z}_{\ge 0}$. The integers $n_\beta$ are labeled by
the class $\beta$ and are the genus zero Gopakumar-Vafa invariants and
the poly-logarithm $\text{Li}_3(x)$ is defined as
$\text{Li}_d(x)=\sum_{n>0} \frac{x^n}{n^d}$ for an integer $d$.

In practice  the prepotential and the periods, that follow from
\eqref{eq:PiXF}, are computed on the B-model side of $M_3$. There they
are solutions to the Picard-Fuchs (PF) differential system associated to the
family of CY three-folds $M_3$ obtained by variation of its
complex structure moduli. We refer to Appendix \ref{app:Review} for the
construction of the PF system that is straightforward for
concrete toric examples. The knowledge of the coefficients in~\eqref{eq:classic_terms_threefold2} is necessary to identify the
linear combination of solutions to the PF system that
is the prepotential $F$ and for which the expression in \eqref{eq:PiXF}
yields periods on $M_3$ in an integral symplectic basis. The instanton
corrections of the A-model side are then readily extracted from
this solution of the PF system in combination with the mirror map
\eqref{eq:t^iIIA}, which we compute using the Mathematica package Instanton
\cite{Albrecht}.

\subsection{Calculations on explicit CY examples}
\label{sec:example1}

In this section we choose a concrete CY three-fold. We explicitly compute  the K\"ahler potential
and flux superpotential from the corresponding prepotential $F$. Before delving
into the details of these computations, we discuss a
general strategy to reduce the typically large complex structure moduli space to a sub-sector of a technically
feasible size.

\subsubsection{Reducing the complex structure moduli by discrete symmetries}
\label{sec:ReducingCSM}

As discussed before, a typical phenomenologically interesting CY three-fold features a small number
of K\"ahler moduli and a large number of complex structure moduli.
Since it is technically unfeasible to obtain the full period vector for
geometries with such a large number of complex structure moduli, we will fix most of the $u$-moduli
by a general principle, i.e.~without doing a
computation. A mechanism to achieve
this was suggested and applied in
\cite{hep-th/0312104,hep-th/0404257,1208.3208}. The basic idea is to
consider a CY three-fold\footnote{In order to avoid confusion, we denote the CY three-fold and its mirror considered here and in
concrete examples by $X_3$, $\tilde{X}_3$.} $X_3$ which admits subsets of its complex
structure
moduli space that are invariant under the action of an appropriate discrete symmetry group.\footnote{As a necessary condition, this
symmetry group has to preserve the CY-condition.}
The symmetry group also acts on the periods of $X_3$.
Switching on fluxes along the invariant three-forms allows to dynamically set the non-invariant complex
structure moduli to their fixed point under the group, e.g.~zero, by solving the F-term conditions as in~\eqref{eq:F-term}.

In the following, we provide a systematic proof of this
argument.\footnote{Arguments similar to the following have been carried out independently by A. Klemm.}
Let us consider a finitely generated symmetry group $G$, as we
will do in the rest of this work.
Consider a group element $g$ in $G$ that acts non-trivially on a single modulus $u^s$, for simplicity, while leaving
all other moduli $u^i$, $i\neq s$ invariant. Denoting the group action by $g\cdot$, $G$-invariant periods and
the K\"ahler potential $K$ obey
\beq \label{eq:invarianceG}
	\Pi(g\cdot u^s)=e^{f(u^s)}\Pi(u^s)\,,\qquad K(g\cdot u^s)=K(u^s)-2\text{Re}(f(u^s))\,,
\eeq
where we suppressed the dependence on $u^i$ for convenience.
Here we demand no strict invariance but allow for a non-trivial gauge transformation $e^f$,
with an appropriate holomorphic function $f(u^s)$. We differentiate both equations in \eqref{eq:invarianceG}
with respect to $u^s$ to obtain the following elementary relations:
\beq \label{eq:derivs}
 g  (\partial_{u^s}\Pi)(g\cdot x)=e^{f(x)}[(\partial_{u^s} \Pi)(x)+  (\partial_{u^s}f)(x)\Pi(x)]\,,\quad
 g (\partial_{u^s}K)(g\cdot x)=(\partial_{u^s}K)(x)-(\partial_{u^s}f)(x)\,,
\eeq
where $x$ is any value for the variable $u^s$.
Equipped with these equations it is straightforward to show for the covariant derivative that
\beq \label{eq:relationDPi}
	g(D_{u^s}\Pi)(g\cdot x)=e^{f(x)} (D_{u^s}\Pi)(x)\,.
\eeq
Thus, we infer that $(D_{u^s} \Pi)(\hat{x})=0$
at every fixed point $\hat{x}$ of the action by $g$, i.e.~at points
with $g\cdot \hat{x} =\hat{x}$, as claimed. This is clear since $g\neq1$, which is the case as we are assuming that $g$ acts
non-trivially on
$u^s$, and since $e^{f(\hat{x})}=1$, as follows from \eqref{eq:invarianceG} at $u^s=\hat{x}$.\footnote{$e^{f(\hat{x})}=1$ holds whenever
the period $\Pi(\hat{x})\neq 0$. If $\Pi(\hat{x})= 0$, then \eqref{eq:derivs} still implies $(D_{u^s}\Pi)(\hat{x})=0$ for
$g\neq e^{f(\hat{x})}$. In addition, if the invariant subset under $G$ of the complex structure moduli space is not only a point, there
always has to be a different invariant period with $\Pi'(\hat{x})\neq 0$ which we can consider.}

In summary, this argument shows that a solution to $D_{u^s}W_{\text{flux}}=0$ for a flux superpotential $W_{\text{flux}}$ with
fluxes switched on only along invariant three-cycles is given by going to the fixed set under the action
of $G$. In other words, for these fluxes the complex structure moduli of $X_3$ can be dynamically driven to this $G$-symmetric locus\footnote{There might be other solutions to the F-term equations for the  $u^s$, that we do not consider here.}.  Furthermore, the invariant
periods, restricted to the $G$-fixed subspace of complex structure moduli, are closed under
the PF operators also restricted to this subspace, since they commute with the group $G$.
The complex structure moduli space obtained in this way agrees with the one of the CY-quotient $X_3/G$.

Concrete examples of such symmetry groups are the Greene-Plesser (GP) orbifold groups
$\Gamma$ of the GP construction~\cite{Greene:1990ud}. This construction
provides mirror pairs of CY three-folds formed by
the three-fold $X_3$
and its quotient $\tilde{X}_3=X_3/\Gamma$ with all
fixed points resolved.
The complex structure moduli space of $\tilde{X}_3$ is embedded into the
complex structure moduli space of $X_3$ as the $\Gamma$-invariant subspace and
all of the above requirements on $G$ are met by $G\equiv \Gamma$. In particular, the invariant
periods restricted to the invariant subspace are simply the periods of the mirror $\tilde{X}_3$.

In the following, we will consider a phenomenologically appealing CY three-fold $X_3$ with four K\"ahler moduli and with
mirror symmetry realised by the GP construction.
We then compute the periods of its quotient CY three-fold $M_3\equiv \tilde{X}_3=X_3/\Gamma$ (cf.~the notation
of Section \ref{sec:periods}) from its PF-system that depends on
only four complex structure moduli.
We stabilise all non-invariant complex structure moduli of $X_3$  at their fixed
point under the GP-group $\Gamma$ by the above mechanism. The remaining periods agree with those of the mirror $\tilde{X}_3$.
Then, stabilisation of complex structure moduli of $X_3$ at the $\Gamma$-symmetric point reduces to stabilising explicitly the remaining
invariant $u$-moduli.

\subsubsection{An explicit example}
\label{sec:example}

The example we consider is a mirror pair of CY hypersurfaces $(X_3,\tilde{X}_3)$ in four-dimensional
toric varieties denoted $\mathbb{P}_{\Delta}$ and $\mathbb{P}_{\tilde{\Delta}}$.
These toric varieties are specified by their reflexive polytopes that we denote by $\Delta_4^{X}$ and $\Delta_{4}^{\tilde{X}}$ for its dual.
The vertices of these polytopes are given by the rows of the following matrices:
\beq
	\Delta_4^{X} =
	\left(
	\begin{array}{r|rrrr||c}
		z_1&1& 1& 0&-1& D_1\\
		z_2&1& 1&-1& 0& D_2\\
		z_3&1& 0& 0& 0& D_3\\
		z_4&1& 0&-1& 1& D_4\\
		z_5&0& 1& 0& 0& D_5\\
		z_6&0& 0& 1& 0& D_6\\
        z_7&0& 0& 0& 1& D_7\\
       z_8&-1&-1& 0& 0& D_8
	\end{array}
	\right),\qquad
	\Delta_4^{\tilde{X}} =
	\left(
	\begin{array}{rrrr||c}
	-1 & -1& -1& -1& \tilde{D}_1\\
	 2 &-1& -1&  2& \tilde{D}_2\\
    2 &-1&  2&  2& \tilde{D}_3\\
	-1&  2& -1&  2& \tilde{D}_4\\
	-1&  2&  2&  2& \tilde{D}_5\\
	 2& -1& -1& -1& \tilde{D}_6\\
	 2& -1&  2& -1& \tilde{D}_7\\
	-1&  2& -1& -1 & \tilde{D}_8
\end{array}
	\right)\,.
	\label{eq:delta4ex1}
\eeq

Here we have introduced the toric divisors $D_i=\{z_i=0\}$ (and $\tilde{D}_i$)
corresponding to the vertices in $\Delta_{4}^{X}$ (and $\Delta_{4}^{\tilde{X}}$) in the last column.
Note that the polytopes $\Delta_4^{X}$ and $\Delta_4^{\tilde{X}}$ are
congruent. Thus the corresponding toric varieties $\mathds{P}_{\Delta}$ and $\mathds{P}_{\tilde{\Delta}}$
differ only by the action of an orbifold group $\Gamma=\mathbb{Z}_3^3$, and the mirror construction
agrees with the GP orbifold construction.

We want to calculate the periods on the complex structure moduli
space of the CY three-fold $X_3$. Its Hodge data and Euler number read
\beq \label{eq:HodgeEuler}
h^{1,1}(X_3)=4 \, 
\,,\quad h^{2,1}(X_3)=70 \, 
\,,\quad \chi(X_3)=-132\,.
\eeq
The subspace that is invariant under the GP orbifold group is only four-dimensional and the
invariant periods at this $\Gamma$-symmetric point precisely agree with the periods
of the mirror $\tilde{X}_3$. By application of the argument of Section~\ref{sec:ReducingCSM} we
reduce the $u$-moduli of $X_3$ to this invariant subspace, which agrees
with the complex structure moduli space of $\tilde{X}_3$. Thus, we present in the remainder
of this section the computation of the ten periods on $\tilde{X}_3$.

First, we note that there are 16 different star-triangulations of $\Delta_4^{X}$ that give rise to eight
different CY phases on $X_3$, some of which correspond to multiple phases of the toric variety and others only to one.
This can be checked by calculating and comparing the triple
intersections on $X_3$ for all these 16 toric phases. The triangulations, the intersections and the following calculations were done by means of PALP \cite{Kreuzer:2002uu,Braun:2012vh} and Sage with the toric geometry package \cite{BraunNovoseltsev:toric_variety}. The Mori cone of a given CY phase is calculated as the intersection of the Mori
cones of the corresponding phases of the ambient space, which are related by flops away from the CY hypersurface.
For simplicity, we choose a CY phase which arises from only one phase of the ambient toric variety and  has a simplicial
Mori cone. It is generated by the $\ell^{(i)}$-vectors
\bea \label{eq:ellsZ3}
	&\ell^{(1)}=(-1, 1, 1, -1, 0, 0, 0, 0)\,,&\qquad \ell^{(2)}=
	(0, -1 , 0, 1, 1, 0, -1, 0)\,,\\
&\ell^{(3)}=(0, 0, -1, 1, 0, 1, -1, 0)\,,&\qquad
\ell^{(4)}=(1, 0, 0, 0, 0, 0, 1, 1)\, .\nn
\eea
The PF-system for $\tilde{X}_3$ will be constructed using \eqref{eq:pfo} from these $\ell^{(i)}$-vectors.
The Stanley-Reissner ideal in terms of the coordinates $z_i$ in \eqref{eq:delta4ex1} in this phase reads
\beq
	\textmd{SR-ideal}:\,\,\{z_2 z_3, z_2 z_6, z_4 z_5, z_4 z_6, z_3 z_5, z_1 z_7 z_8\}\,.
\eeq
By duality between curves and divisor:
\beq
	K_i\cdot \ell^{(j)}=\delta^{j}_i\,,
\eeq
the $\ell^{(i)}$-vectors \eqref{eq:ellsZ3} also fix the generators $K_i$ of the K\"ahler cone. In the chosen
triangulation, the K\"ahler cone is spanned by the generators
\beq \label{eq:Kaehlerconeex1}
	K_1=D_2+D_3+D_4\,,\quad K_2=D_5\,,\quad K_3=D_2+D_4\,,\quad
	K_4=D_1+D_3+D_6\,,
\eeq
in terms of the toric divisors $D_i$ introduced in \eqref{eq:delta4ex1}.

As a cross-check, we confirm explicitly that there exists a CY phase of the mirror $\tilde{X}_3$ that contains
this chosen topological phase of $X_3$ as a sub-sector of its K\"ahler
moduli space. This confirms, invoking mirror symmetry, that the complex structure moduli space
of $\tilde{X}_3$ is indeed a sub-sector of the complex structure moduli space of $X_3$, as guaranteed by
the GP construction.

In the basis \eqref{eq:Kaehlerconeex1} of the K\"ahler cone
on $X_3$ we calculate the triple intersections $\mathcal{K}_{ijk}$.
We summarise these intersections on $X_3$ in a formal polynomial
$\mathcal{C}_0=\frac{1}{6!}\mathcal{K}_{ijk}J_iJ_jJ_k$ in the dual (1,1)-forms $J_i$ that reads:\footnote{We have not
chosen the intersection form $I_3$ here as $\mathcal{C}_0$ is directly related to the cubic terms in $F$.}
\beq \label{eq:C0ex1}
	\mathcal{C}_0=\frac{3}{2} J_1^2 J_4 + 3 J_1 J_2 J_4 + 3 J_1 J_3 J_4 +
 3 J_2 J_3 J_4 + \frac{9}{2} J_1 J_4^2 + 3 J_2 J_4^2 +
 3 J_3 J_4^2 + \frac{5}{2} J_4^3\,.
\eeq
Analogously we calculate the
other classical intersections in \eqref{eq:classic_terms_threefold2} to be
\bea \label{eq:subleadingtermsex1}
	\cK_{14}= -\frac{9}{2}\,,\,\, \cK_{24}=-3\,,\,\,\cK_{34}=-3\,,\,\,\,\,
	 \cK_{44}= -\frac{15}{2}\,,\,\,\ \cK_{41}= -\frac{3}{2}\,, \\
  \cK_j J_j = \frac{3}{2}J_1+J_2+J_3+\frac{33}{12}J_4\ , \qquad
  \cK_0= -\frac{\zeta(3)}{(2 \pi i)^3}132\,,
\eea
with all other $\cK_{ij}=0$. We note that, as before, these relations hold up to integers
and the $\cK_{ij}$ are symmetric modulo integers.

Finally we obtain the prepotential \eqref{eq:pre_largeV} on $\tilde{X}_3$ using the
intersections \eqref{eq:C0ex1} and \eqref{eq:subleadingtermsex1}:
\bea
F\!&\!=\!&\!\!-\frac{3}{2} (u^1)^2 u^4 - 3 u^1 u^2 u^4 - 3 u^1 u^3 u^4 -
 3 u^2 u^3 u^4 - \frac{9}{2} u^1 (u^4)^2 - 3 u^2 (u^4)^2 - 3 u^3 (u^4)^2  \nn \\
 &-&\frac{5}{2} (u^4)^3+3 u^1 u^4+\frac{3}{2} u^2 u^4+\frac{3}{2}u^3u^4+\frac{15}{4}(u^4)^2
 +\frac{3}{2}u^1+u^2+u^3+\frac{33}{12}u^4-i\zeta(3)\frac{33}{4\pi^3}\nn\\
 &+&  \sum_{\beta} n_\beta^0\,\text{Li}_3(q^\beta)\,,
\label{eq:prepotentialexample}	
\eea
where $\beta=(d_1,d_2,d_3,d_4)$ in the basis $\beta^i$ of effective curves of $H_2(X_3,\mathbf{Z})$
corresponding to the charge vectors \eqref{eq:ellsZ3} and $q^\beta=e^{2\pi id_iu^i}$.
We also introduced the flat coordinates $u^i$ on $\tilde{X}_3$, that are identified with the K\"ahler moduli of
$X_3$ via the mirror map \eqref{eq:t^iIIA}. The instanton
corrections in \eqref{eq:prepotentialexample} are determined by solving the
PF equations on $\tilde{X}_3$ that follow from \eqref{eq:ellsZ3}. First we
calculated explicitly the mirror map \eqref{eq:t^iIIA} and then identified
the prepotential $F$ among the solutions of the PF system by
matching the above classical terms \eqref{eq:C0ex1} and \eqref{eq:subleadingtermsex1}.
Then the worldsheet instanton corrections
at large volume are uniquely determined and read off by comparing to
the general large-volume expansion of $F$ in \eqref{eq:pre_largeV}.\footnote{The explicit instanton corrections can be found in the accompanying material of this paper on the arXiv.}

Equipped with \eqref{eq:prepotentialexample} we readily compute the full period
vector according to \eqref{eq:periodExp} and the K\"ahler potential $K$ as well as the flux superpotential
for arbitrary flux choices from \eqref{eq:KaehlerpotCS}. We refer to Appendix~\ref{app:Review}, in particular
\eqref{eq:periodVectorCY3LV} and \eqref{eq:flux_ori}, for more explicit formulae of
$\Pi$ and $W_{\text{flux}}$ in terms of the prepotential $F$. The explicit stabilisation of dilaton and complex structure moduli on $X_3$ at the $\Gamma$-invariant locus is performed in Section~\ref{sec:stabilisation}.


\section{Explicit model}
\label{sec:ExplicitModel}

On the three-fold $X_3$ introduced in the previous section, we now construct an explicit chiral D-brane setup which satisfies all stringy consistency requirements, allows for a chiral visible sector and gives rise to a hidden sector suitable for K\"ahler moduli stabilisation. For this construction of a chiral global model we follow the same philosophy as in~\cite{Cicoli:2012vw,1304.0022}.

In section~\ref{sec:geometric} we study the properties of $X_3$ and find that it has three dP$_6$ divisors, and two of these divisors can be exchanged by an orientifold involution with O3/O7 planes. On these exchangeable dP$_6$ divisors, which will be stabilised at zero size, we place a visible sector setup with fractional D3-branes. In section~\ref{sec:ModelBuilding} we analyse the gauge theory which arises in this sector. For simplicity we concentrate on a setup with no flavour D7-branes. Equipped with this visible sector, we then construct a minimal hidden sector D-brane setup in section~\ref{sec:hiddendbranes} compatible with all stringy consistency conditions and suitable for moduli stabilisation.

\subsection{Geometric setup and orientifold involution}
\label{sec:geometric}

The weight matrix for $\Delta_4^{X}$ given in
\eqref{eq:delta4ex1} reads as follows:\footnote{The
triangulation used here is the one relevant for the construction of
the CY orientifold of $X_3$. It differs from the one in
\eqref{eq:ellsZ3}. The periods on $X_3$ are, however, unaffected by
this change of triangulation.}
\begin{equation}
\begin{array}{|c|c|c|c|c|c|c|c||c|}
\hline z_1 & z_2 & z_3 & z_4 & z_5 & z_6 & z_7 & z_8 & D_{eq_{{Z}}} \tabularnewline \hline \hline
1 &  0 &  0 &  1 &  1 &  1 &  0 &  2 & 6\tabularnewline\hline
1 &  0 &  0 &  0 &  0 &  0 &  1 &  1 & 3\tabularnewline\hline
0 &  1 &  0 &  0 &  0 &  1 &  0 &  1 & 3\tabularnewline\hline
0 &  0 &  1 &  0 &  1 &  0 &  0 &  1 & 3\tabularnewline\hline
\end{array}
\label{eq:ex1:weightm}\, .
\end{equation}
The Stanley-Reisner ideal for the chosen triangulation of the ambient
four-fold is:\footnote{Note that out of the 16 triangulations of
$\Delta_4^{X}$ there are three more which give the same intersection
ring on the CY hypersurface.}
\begin{equation} \label{eq:SRideal}
\textmd{SR-ideal}:\,\,\{z_2\,z_3,\, z_3\,z_5,\, z_2\,z_6,\, z_2\,z_7,\, z_3\,z_7,\, z_4\,z_5\,z_6,\, z_1\,z_7\,z_8,\, z_1\,z_4\,z_5\,z_8,\, z_1\,z_4\,z_6\,z_8\}
\end{equation}
Each column in \eqref{eq:ex1:weightm} denotes the scaling behaviour of each homogeneous coordinate $z_i$ under the four $\mathbb{C}^*$-actions of the ambient fourfold. The last column of the table determines  the CY hypersurface
by fixing its defining equation to be of the form
\begin{equation}
   X_3\,: \qquad   P_{6,3,3,3}(z_i) = 0\,,
\end{equation}
with $P_{6,3,3,3}(z_i)$ a polynomial of degrees $(6,3,3,3)$ with respect to the $(\mathbb{C}^*)^4$-action of \eqref{eq:ex1:weightm}.

The CY three-fold $X_3$ has three dP$_6$ divisors at $z_2=0$, $z_3=0$ and $z_7=0$.
We can take these three divisors as part of an integral `diagonal' basis of $H^{1,1}(X_3)$:
\be
\cD_{q_1} = D_2\,,\qquad \cD_{q_2} = D_3\,,\qquad \cD_{b} = D_1+D_2+D_3\,,\qquad \cD_{s} = D_7\,,
\label{basis}
\ee
i.e.~a basis for which the intersection form $I_3$ has no cross-terms:
\be
 I_3= 3\cD_{q_1}^3+3\cD_{q_2}^3+3\cD_b^3+3\cD_s^3 \,.
\ee
Expanding the K\"ahler form $J$ in the basis (\ref{basis}) as\footnote{By abuse of notation,
we denote the $(1,1)$-forms dual to the divisors \eqref{basis} by the same symbol.} $$J= t_b \cD_b + t_{q_1}\cD_{q_1} + t_{q_2}\cD_{q_2}+t_s\cD_s\,,$$
the volumes of the four divisors take the form:
\be
{\rm Vol}(\cD_i)\equiv \tau_i =\frac 12 \int_{\cD_i} J\wedge J = \frac 32\, t_i^2\,,
\ee
while the CY volume becomes of strong `Swiss-cheese' type:
\be
{\rm Vol}(X_3) \equiv \vo = \frac 16 \int_{X_3} J\wedge J\wedge J= \sum_i \frac{t_i^3}{2}
=\frac{1}{3}\sqrt{\frac{2}{3}}\left[\tau_b^{3/2} -  \left( \tau_{q_1}^{3/2} + \tau_{q_2}^{3/2} + \tau_s^{3/2}\right)\right]\,.
\label{vol}
\ee
Note that the minus sign in (\ref{vol}) is due to the fact that the K\"ahler cone conditions for the three dP$_6$ divisors
are
\begin{equation}\label{eq:kaehler_cone-partI}
t_{q_1}<0,\quad t_{q_2}<0,\quad t_s<0\,.
\end{equation}
The additional conditions involving $t_b$ read as follows:
\begin{equation}\label{eq:kaehler_cone-partII}
t_b+t_s+t_{q_1}>0, \quad t_b+t_s+t_{q_2}>0,\quad t_b+t_{q_1}+t_{q_2}>0\,.
\end{equation}
This K\"ahler cone is obtained by the union of the four K\"ahler cones corresponding to the triangulations of $\Delta_4^{X}$ which give the
same toric intersection numbers on the CY hypersurface. Since this cone is identical to the cone generated by the curves coming from
intersections of the hypersurface equation with two toric divisors, we conclude that \eqref{eq:kaehler_cone-partI} and
\eqref{eq:kaehler_cone-partII} span indeed the K\"ahler cone of the CY hypersurface \cite{Berglund:1995gd}.

The involution which exchanges the two dP$_6$ divisors $\cD_{q_1}$ and $\cD_{q_2}$ is given by:
\begin{equation}\label{eq:involution-model1}
 \sigma:\quad (z_2\,,\,z_5)\leftrightarrow(z_3\,,\,z_6)\,.
\end{equation}
It is readily checked that it is a symmetry of the weight matrix \eqref{eq:ex1:weightm}.
In the following we take the limit in K\"ahler moduli space in which these two dP$_6$'s shrink to zero size. The
two singularities generated in this way are exchanged under the orientifold involution, as required. The remaining
dP$_6$ surfaces at $z_7=0$ ($\cD_s$) is kept at finite size. Note that this desired structure of the K\"ahler moduli arises in the process of moduli stabilisation discussed in Section~\ref{sec:stabilisation}.

The equation defining the CY hypersurface must be symmetric under the orientifold involution. This gives the following restricted equation
\begin{equation} \label{eq:Psym}
P_{6,3,3,3}^{\rm sym}(z_i) \equiv  P_{6,3,3,3}(z_1,z_2,z_3,z_4,z_5,z_6,z_7,z_8) + P_{6,3,3,3}(z_1,z_3,z_2,z_4,z_6,z_5,z_7,z_8) =0\:.
\end{equation}

The fixed point set of the involution \eqref{eq:involution-model1} can be determined by considering the (anti-) invariant combinations $y_i$ of the homogeneous coordinates $z_2$, $z_3$, $z_5$ and $z_5$:
\begin{equation}
y_2\equiv z_2z_3, \qquad y_3\equiv z_5z_6, \qquad
y_5\equiv z_2z_5+z_3z_6, \qquad y_6\equiv z_2z_5-z_3z_6\,.
\end{equation}
They are all invariant under the orientifold involution, except $y_6$ which transforms as $y_6\mapsto - y_6$.
Hence, the hypersurface $y_6=0$ belongs to the fixed locus of \eqref{eq:involution-model1}. Its intersection with $X_3$ gives the following
O7-plane:
\begin{equation} \label{eq:O7plane}
\text{O7}:\quad z_2\,z_5-z_3\,z_6=0\, \qquad \text{with} \qquad \cD_{O7}=\cD_b-\cD_s\,,
\end{equation}
and $\chi(\cD_{\text{O7}})=36$.
There are also other points in the fixed point set. In fact, we can compensate the minus one factor of $y_6$ by using the scaling relations of the
toric coordinates \eqref{eq:ex1:weightm}. Taking this into account,
we find a codimension-three and a codimension-four fixed locus in the ambient space. The codimension-three locus is
\begin{equation}\label{codim3fixedloc}
 \{z_1 = 0,\,\,\,z_4 =0,\,\,\, y_5\equiv z_2\,z_5 +z_3\, z_6=0\}\:,
\end{equation}
that, once intersected with $X_3$, gives three O3-planes.\footnote{If we intersect \eqref{codim3fixedloc} with the hypersurface equation of $X_3$, it reduces to a degree three polynomial which has three solutions each of which is an O3-plane.} The codimension-four locus is
\begin{equation}\label{codim4fixedloc}
\{ z_4= 0,\,\,\, y_5\equiv z_2\, z_5 +z_3\, z_6=0,\,\,\, z_7= 0,\,\,\,z_8= 0\} \,.
\end{equation}
This is one point in the ambient space which lies on the symmetric CY hypersurface given by $P_{6,3,3,3}^{\rm sym}(z_i)=0$ and, therefore, gives one
O3-plane.\footnote{The relations \eqref{codim4fixedloc} are four equations in the ambient four-fold that automatically solve the orientifold
symmetric equation of the CY three-fold $P_{6,3,3,3}^{\rm sym}(z_i)=0$.} Thus, in total we have one O7-plane and four O3-planes. For the
details of the determination of the O3- and O7-planes we refer to \cite{Mayrhofer2010,Cicoli:2011it,
Cicoli:2012vw}, where the fixed-point locus of a
holomorphic involution for a different CY three-fold has been determined using identical techniques.

After having derived the fixed point set of our involution, we can calculate, by means of Lefschetz fixed point theorem \cite{Shanahan1978},\footnote{For applications of this theorem in the physics literature see for instance \cite{Blumenhagen:2010ja,Distler:1987ee,Brunner:2003zm}.} the number of complex structure deformations after imposing invariance under
\eqref{eq:involution-model1}, i.e.\ $h^{1,2}_-$.
The fixed point theorem states that the differences between the alternating sum over the odd and the even Betti numbers is just the Euler characteristics of the fixed point set of the involution:
\begin{equation}\label{eq:lefschetz-fixed-point}
 \sum_i (-1)^i(b_+^i-b_-^i)=\chi(O_\sigma)\qquad\textmd{with} \qquad b_\pm^i=\sum_{p+q=i}h_\pm^{p,q}\,.
\end{equation}
Our fixed point set $O_\sigma$ is given by one O7-plane with $\chi(\textmd{O7})=36$ and four O3-planes with each $\chi(\textmd{O3})=1$ which gives in total 40 for the Euler number of $O_\sigma$.
In addition, we know that $h^{1,1}_-=1$, $h^{1,1}_+=3$, $h^{2,1}=70$, $h^{3,0}=h^{3,0}_-=1$ ($\Omega\in H^{3,0}_-(X_3)$) and that we are dealing with a proper CY three-fold, i.e.\ $h^{1,0}=h^{2,0}=0$.
These data together with \eqref{eq:lefschetz-fixed-point} give us $h^{2,1}_-=43$ and $h^{2,1}_+=27$.

Next, we consider in more detail the GP orbifold group $\Gamma$ used for the construction of the mirror
$\tilde{X}_3$. This is essential in order to understand the complex structure moduli space of $X_3$ that is
both invariant under $\Gamma$ and the orientifold involution \eqref{eq:involution-model1}.
The GP group $\Gamma=\mathbb{Z}_3^3$ has the following generators:
\begin{eqnarray}\label{eq:ex1-Z3-action}
\gamma_1\,:\qquad (z_1,z_2,z_3,z_4,z_5,z_6,z_7,z_8) &\mapsto & (z_1,z_2,z_3,z_4,z_5,z_6,e^{2\pi i/3}z_7,e^{4\pi i/3}z_8) \nonumber \\
\gamma_2\,:\qquad (z_1,z_2,z_3,z_4,z_5,z_6,z_7,z_8) &\mapsto & (z_1,z_2,z_3,z_4,e^{2\pi i/3}z_5,z_6,ez_7,e^{4\pi i/3}z_8) \\
\gamma_3\,:\qquad (z_1,z_2,z_3,z_4,z_5,z_6,z_7,z_8) &\mapsto & (z_1,z_2,z_3,z_4,z_5,e^{4\pi i/3}z_6,e^{2\pi i/3}z_7,z_8) \nonumber
\end{eqnarray}
From this, we infer that the orientifold invariant CY equation $P_{6,3,3,3}^{\rm sym}=0$ in \eqref{eq:Psym} takes the following restricted form,
if \eqref{eq:ex1-Z3-action} is imposed to be a symmetry as well:
\begin{equation}\label{eq:reduced-HSE-model1}
\begin{aligned}
P_{6,3,3,3}^\textmd{rd}(z_i)=&
z_1^3 \, z_2^3 \, z_5^3 +
z_1^3 \, z_3^3 \, z_6^3 +z_2^3 \, z_4^3 \, z_5^3 \, z_7^3 +
z_3^3 \, z_4^3 \, z_6^3 \, z_7^3 +
z_5^3 \, z_6^3 \, z_7^3 +
z_8^3 +
\\& +
\rho_1 \, z_2^3 \, z_3^3 \, z_4^6 \, z_7^3 +
\rho_2 \, z_1^3 \, z_2^3 \, z_3^3 \, z_4^3 +
\rho_3 \, z_1 \, z_2 \, z_3 \, z_4 \, z_5 \, z_6 \, z_7 \, z_8    \:.
\end{aligned}
\end{equation}
In this equation, we have fixed completely the reparametrisation invariance. Hence, the remaining undetermined three coefficients are related to
the complex structure moduli that survive both the orientifold and the discrete symmetry projection. Note that before imposing the orientifold
projection we have four of them, as expected from
Section \ref{sec:example}.

\subsection{Visible sector D-branes}
\label{sec:ModelBuilding}

Let us discuss the field theory on the D3-branes at the dP$_6$ singularity of $X_3/\sigma$.
We construct the visible sector of our model by placing $N_{D3}$ D3-branes on top of the dP$_6$ singularity at $z_2=0$ (plus their $N_{D3}$
images at the singularity at $z_3=0$) of $X_3$.
As explained above, these singularities are generated by shrinking the two dP$_6$ divisors, $\cD_{q_1}$ and $\cD_{q_1}$, exchanged by
the orientifold involution.
We do not consider additional flavour D7-branes for simplicity, because their presence would also affect the hidden D7-brane setup described
in the next section.\footnote{If the local gauge theory construction involves flavour D7-branes, one has also to add some D7-branes which
can play their r\^ole in the global configuration. In this case, the D7-tadpole induced by the O7-plane should not be completed saturated by the
stack of D7-branes on top of the O7 (see \cite{1201.5379,1304.0022,1304.2771}). In the presence of flavour D7-branes, we might keep $N<4$
D7-branes on top of the O7-plane and the form of the FI-term in (\ref{FIb}) would still be the same, only the rank of
the $U(N)$ gauge group would differ.}

\begin{figure}[t]\begin{center}
 \includegraphics[width=0.3\textwidth]{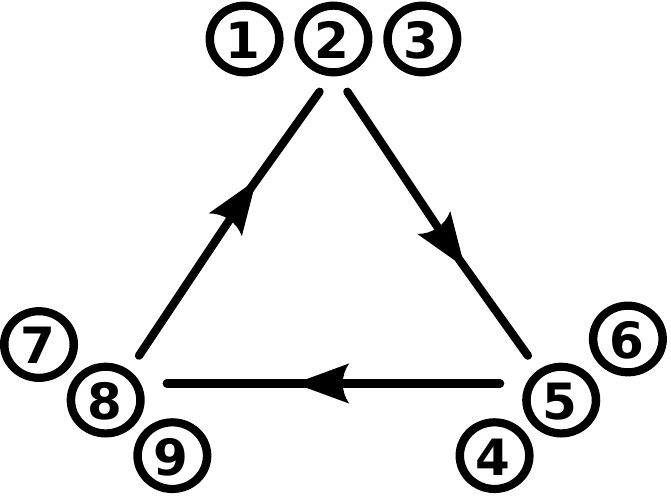}\hspace{0.3cm}
 \includegraphics[width=0.18\textwidth]{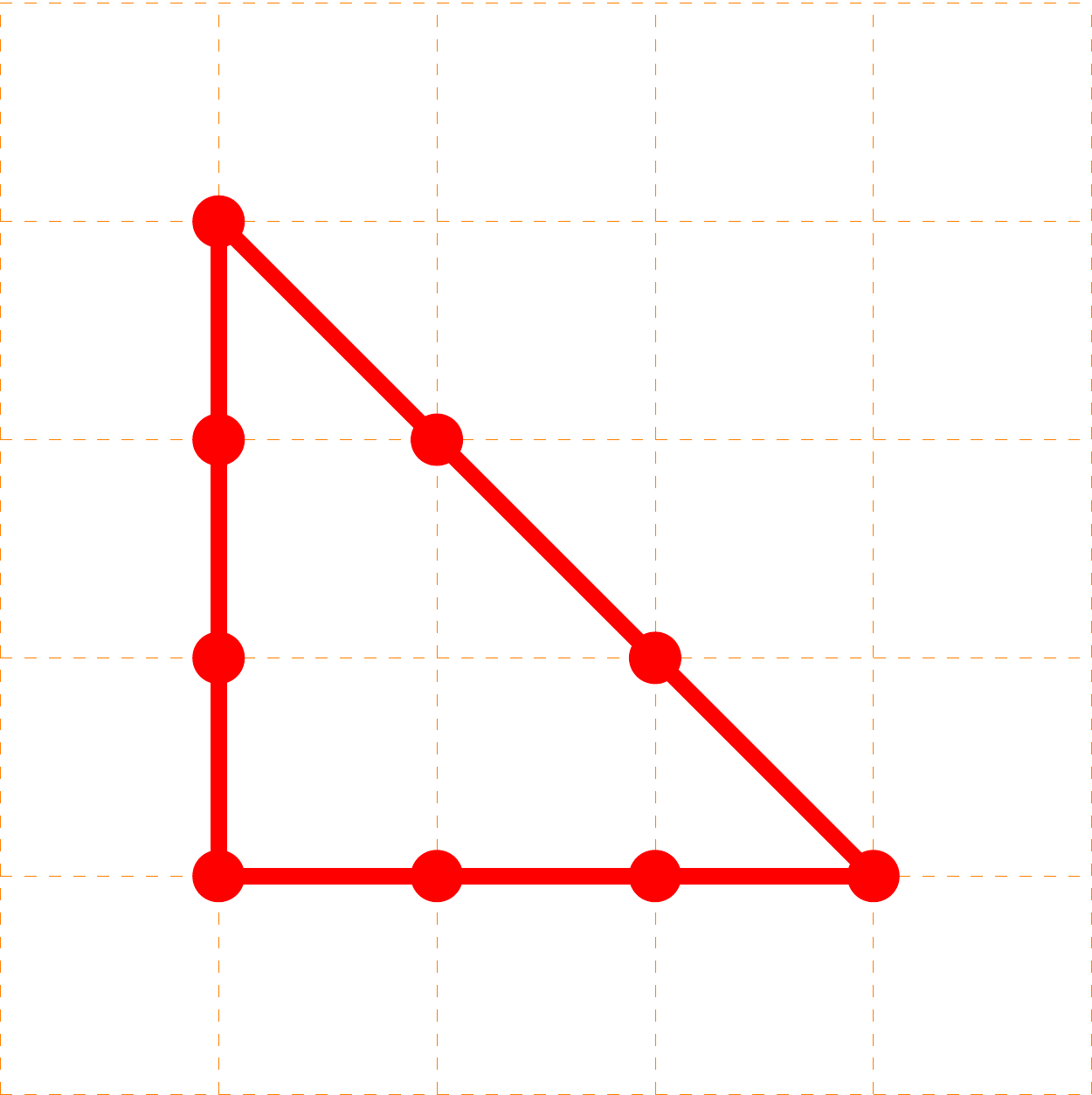}\hspace{0.5cm}
  \includegraphics[width=0.25\textwidth]{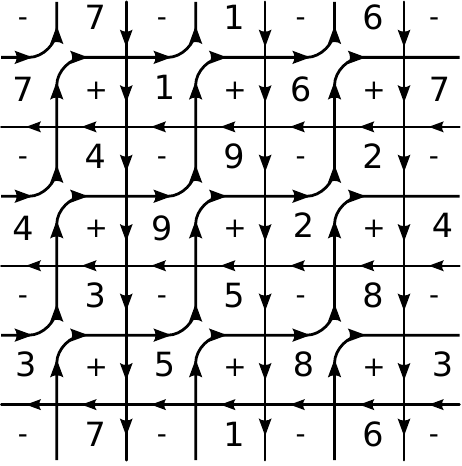}
\captionof{figure}{\footnotesize{From left to right, the quiver diagram, the toric diagram and the dimer diagram for the
$\mathbb{C}^3/\mathbb{Z}_3\times\mathbb{Z}_3$ orbifold singularity.}\label{fig:c3z3z3}}
\end{center}\end{figure}

\begin{figure}[t]\begin{center}
 \includegraphics[width=0.3\textwidth]{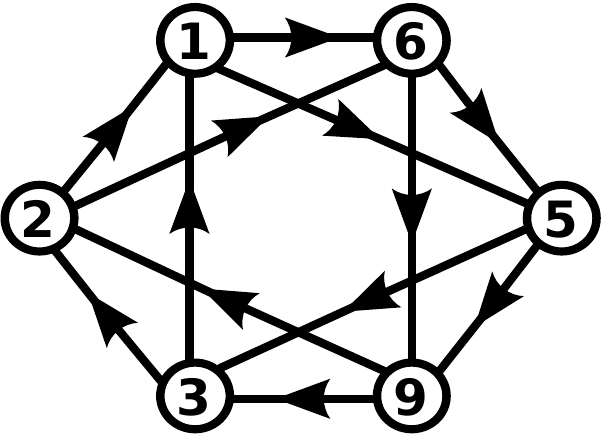}\hspace{1cm}
  \includegraphics[width=0.25\textwidth]{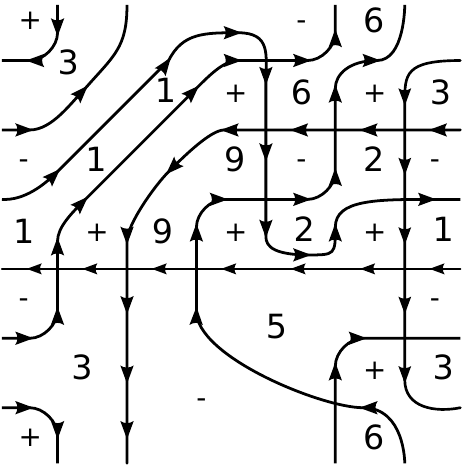}
\captionof{figure}{\footnotesize{After assigning VEVs for the fields $X_{14},$ $Y_{58}$, $Z_{73}$ the matter content shown in the quiver diagram on the left side remains. The resulting dimer after the higgsing is shown on the right side. The remaining gauge theory is that of dP$_3$.}\label{fig:dp3s}}
\end{center}\end{figure}

Let us briefly comment on how this dP$_6$ singularity can be the basis for model building. As pointed out in~\cite{Wijnholt:2002qz} there
exists a gauge theory description of the dP$_6$ singularity which locally is a $\mathbb{C}^3/\mathbb{Z}_3\times \mathbb{Z}_3$ orbifold
singularity. We use this gauge theory description in terms of the $\mathbb{C}^3/\mathbb{Z}_3\times \mathbb{Z}_3$ orbifold as the starting
point for our model building discussion. Using dimer techniques, we find the dimer and quiver diagram as shown in Figure~\ref{fig:c3z3z3}.
Nodes in the quiver correspond to $U(N_i)$ gauge theories, arrows to bi-fundamental fields. The dimer can be obtained from the toric diagram
using various inverse algorithms (e.g.~\cite{Gulotta:2008ef,Hanany:2005ss,Yamazaki:2008bt}). Numbered faces in the dimer correspond to
gauge groups, common nodes of faces to bi-fundamental matter, and the ``$\pm$'' faces correspond to superpotential terms. From the dimer
in Figure~\ref{fig:c3z3z3} we can then read off the
following superpotential
\begin{eqnarray}
\nonumber W&=&X_{14} Y_{47} Z_{71}-X_{14} Y_{49} Z_{91}-X_{15} Y_{57} Z_{71}+X_{15} Y_{58} Z_{81}-X_{16} Y_{68} Z_{81}\\
\nonumber  &&+X_{16} Y_{69} Z_{91}-X_{24} Y_{47} Z_{72}+X_{24} Y_{48} Z_{82}-X_{25} Y_{58} Z_{82}+X_{25} Y_{59} Z_{92}\\ \nonumber
&&+X_{26} Y_{67} Z_{72}-X_{26} Y_{69} Z_{92}-X_{34} Y_{48} Z_{83}+X_{34} Y_{49} Z_{93}+X_{35} Y_{57} Z_{73}\\ &&-X_{35} Y_{59} Z_{93}-X_{36}
Y_{67} Z_{73}+X_{36} Y_{68} Z_{83}\,.
\label{Wquiver}
\end{eqnarray}
The fields $X_{ij}$, $Y_{kl}$, $Z_{mn}$ are labelled by different letters for each of the three different arrows in
the quiver in Figure~\ref{fig:c3z3z3} and the indices in their labelling denote the transformation behaviour as bi-fundamental fields
$(N_i,\bar{N}_j)$.
This gauge theory can be related to phenomenologically interesting gauge theories via higgsing. In particular by assigning a VEV to $X_{14},$ $Y_{58},$ and $Z_{73},$ we find the following superpotential after integrating out all states with superpotential mass terms:
\begin{eqnarray}
\nonumber W&=&-X_{16} X_{21} Y_{59} Y_{65} Z_{32} Z_{93}+X_{16} X_{31} Y_{69} Z_{93}+X_{21} Y_{15} Y_{59} Z_{92}\\ &&+X_{26} Y_{65} Z_{32} Z_{53}-X_{26} Y_{69} Z_{92}-X_{31} Y_{15} Z_{53}
\end{eqnarray}
This process is also visualised in Figure~\ref{fig:dp3s} which shows the dimer diagram after assigning these VEVs and the associated quiver diagram.

This gauge theory is the gauge theory of dP$_3$ which has proven to be phenomenologically very appealing~\cite{1106.6039}. Models based on the Pati-Salam gauge group have been constructed in~\cite{1106.6039} where it was shown that they can, in the presence of flavour D7-branes, give rise to a breakdown to the SM gauge group, an appealing flavour structure for quarks and leptons, the absence (respectively sufficient suppression) of proton decay and a viable $\mu$ term.



\subsection{Hidden sector D-branes}
\label{sec:hiddendbranes}

In this section we discuss the D-brane setup away from the
dP$_6$-singularity. This includes a discussion
of both the D7-branes necessary for D7 tadpole cancellation as well as possible non-perturbative
effects from D3-brane instantons and gaugino condensates.

\subsubsection*{E3 instantons}

We need to consider D-branes which generate non-vanishing terms in the non-perturbative superpotential. This is relevant in order to fix some
of the K\"ahler moduli as we will see in Section \ref{sec:KaehlerStab}.
The non-perturbative superpotential is generated by E3-instantons and/or gaugino condensation on stacks of D7-branes wrapping four-cycles
in the CY three-fold (see for instance~\cite{Blumenhagen:2009qh} for a review).
In the LVS the leading contribution comes from a shrinkable rigid divisor (e.g.~del Pezzo), if
this can support non-perturbative effects.

In our model based on the CY orientifold $X_3/\sigma$, we do not have D7-branes wrapping the
(orientifold invariant) del Pezzo divisor $\mathcal{D}_s$,
because the O7-plane charge on the divisor \eqref{eq:O7plane} cannot compensate the D7 tadpole of these
D7-branes.
Hence the possible leading non-perturbative effect arises from $O(1)$
E3-instantons wrapping the divisor $\cD_s$ which is mapped to itself
under the orientifold involution $\sigma$. Since the fermionic zero
modes of such instantons are projected out,
whether the E3's contribute to the superpotential
depends only on their world-volume fluxes.
The divisor $\cD_s$ is non-spin and hence Freed-Witten (FW) anomaly
cancellation enforces a non-zero gauge flux \cite{Freed:1999vc}:
\begin{equation}\label{FWcancE3}
 F_{E3} +\frac{c_1(\cD_s)}{2} \,\, \in \,\, H^2(\cD_s,\mathbb{Z}) \:.
\end{equation}
This quantisation implies that the gauge invariant field strength $\cF_{E3}=F_{E3}-i^\ast_{\cD_s} B$ is non-zero, if the pullback $i^\ast_{\cD_s} B$ of the NSNS $B$-field to the world-volume $\cD_s$ of the brane is zero. Hence, an $O(1)$ E3-instanton wrapping such a divisor is not invariant under the orientifold involution and is projected out. \footnote{A two-form $F_{E3}$ satisfying \eqref{FWcancE3} is always even when $\cD_s$ has zero intersection numbers with the odd $h^{1,1}_-$ divisors, as it happens in the considered example.}

If we insisted on having $\iota_{\cD_s}^*B=0$, the leading contribution
to the superpotential arises from a rank-two E3-instanton
\cite{1210.1221}, which we will however not consider in the following.
In fact, in the LVS a leading
non-perturbative contribution to the superpotential from rank-one
E3-instantons is phenomenologically preferred
compared to rank-two E3-instantons.
The volume in this minimum is fixed to be
$\vo\sim |W_0|e^{a\tau_s}$. For a rank-one E3-instanton we have $a=2\pi$, while for a rank-two E3-instanton
$a=4\pi$. Since $\tau_s>1$ and $|W_0|\sim \mathcal{O}(1)$, the second option would fix the volume of the CY at a
too large value, rendering the phenomenologically relevant scale too small. We will comment on this in Section \ref{sec:KaehlerStab}.

Thus we choose a non-zero B-field
\begin{equation}\label{BE3constr}
 i^\ast_{\cD_s}B = \frac{\cD_s}{2} \,,
\end{equation}
so that we can pick a gauge flux $F_{E3}$ such that $\cF_{E3}=0$ \cite{Collinucci:2008sq}. In this case the $O(1)$ E3-instanton is
invariant and survives the orientifold projection.

All other possible E3-instantons wrap divisors that, in the LVS minimum, have a larger size compared to $\cD_s$.
Thus, their contributions to the superpotential are exponentially suppressed and can be consistently
neglected.

\subsubsection*{Hidden D7-branes}

To cancel the D7-tadpole introduced by the O7-plane, we place four D7-branes (plus their images)
on top of the O7-plane locus $\cD_{O7}$ given in \eqref{eq:O7plane}. This generates a hidden $SO(8)$ gauge group.
The number of deformation moduli of such D7-branes is
$h^{0,2}(\cD_{O7})=2$. There are no Wilson-line moduli as
$h^{0,1}(\cD_{O7})=0$.

If the $B$-field
is zero, this group is broken to $U(4)$ by a properly quantised non-zero gauge flux. In fact, FW anomaly
cancellation implies the flux to be quantised such that
\begin{equation}\label{FD7FWquant}
 F_{D7} +\frac{c_1(\cD_{D7})}{2} \,\, \in \,\, H^2(\cD_{D7},\mathbb{Z}) \:.
\end{equation}
Since $\cD_{D7}=\cD_{O7}$ is an odd cycle, the flux is necessary non-zero.\footnote{Alternatively,
if we choose the $B$-field to be (consistently with \eqref{BE3constr}):
\be
B = - \frac{c_1(\cD_{O7})}{2} = \frac{\cD_b}{2} - \frac{\cD_s}{2}\:,
\label{Bchoice1}
\ee
the gauge invariant flux can be set to zero,
$\cF_{D7}=F_{D7}-i^\ast_{\cD_{O7}}B=0$. Thus, the $SO(8)$ gauge group
is unbroken. By switching on a proper rigidifying flux, we
could lift the deformation moduli\cite{Bianchi:2011qh,Bianchi:2012kt,1208.3208}. Hence, for the $B$-field
\eqref{Bchoice1}, we have a pure super
Yang-Mills theory in the hidden sector which undergoes gaugino condensation.
However, as we shall see in Section \ref{sec:KaehlerStab}, this setup would not give rise to any D-term
associated to an anomalous $U(1)$ living on $\cD_{D7}$ which is a necessary ingredient to obtain de Sitter vacua.}

We choose in the following the minimal choice for the $B$-field that
satisfies \eqref{BE3constr}:
\be
B = - \frac{c_1(\cD_s)}{2} = \frac{\cD_s}{2} \,.
\label{Bchoice2}
\ee
We consider a gauge flux $F_{D7}$ on the brane world-volume which is a linear combination of pulled back two-forms from the CY three-fold. Since the intersections of $X_3$, c.f.~\eqref{eq:SRideal}, imply that  $i^\ast_{\cD_{O7}}\cD_{q_1}=i^\ast_{\cD_{O7}}\cD_{q_2}=0$, the
flux can only be a linear combination of $i^\ast_{\cD_{O7}}\cD_b$ and $i^\ast_{\cD_{O7}}\cD_s$ which we still call
$\cD_b$ and $\cD_s$ by abuse of notation.
By taking into account \eqref{FD7FWquant}, \eqref{Bchoice2} and $c_1(\cD_{O7})=-\cD_b+\cD_s$, the generic invariant and quantised flux on the brane is:
\begin{equation}
 \cF_{D7} = F_{D7}-i^\ast_{\cD_{O7}}B
 =  \left(n_{b}+\frac12 \right)\cD_b + \left(n_s-1\right)\cD_{s}  \,,
 \label{eq:fluxD7}
\end{equation}
with $n_{b},n_{s}\in \mathbb{Z}$. On the image stack the flux is the opposite.
Given that the flux along $\cD_b$ cannot be cancelled due to our $B$-field choice,
$\cF_{D7}$ is different from zero.
This has a few consequences:
\begin{itemize}
\item $\cF_{D7}\neq 0$ breaks $SO(8)\to SU(4) \times U(1)$ where the anomalous $U(1)$ factor gives rise to a D-term contribution to the scalar potential.
The corresponding Fayet-Iliopoulos (FI) term reads \cite{Jockers:2005zy}:
\be
\xi_{D7} = \frac{1}{\vo}\int_{\cD_{D7}} J\wedge \cF_{D7}
=\frac{3}{\vo}\left[\left(2 n_b+1\right)\frac{t_b}{2}+\left(1-n_s\right)t_s\right].
\label{FIbig}
\ee

\item $\cF_{D7}\neq 0$ generates massless chiral matter living on the hidden sector D7-branes.
The number of chiral fields in the antisymmetric representation of
$U(4)$ is (see for instance~\cite{Blumenhagen:2006ci}):
\be
I_{U(4)} = \frac 12\,I_{D7-D7'}+ I_{D7-O7}\,,
\label{IU4}
\ee
where:
\be
I_{D7-D7'}=\int_{\cD_{D7}\cap \cD_{D7'}}\left(\cF_{D7}-\cF_{D7'}\right)
= 2\int_{X_3} \cD_{D7} \wedge \cD_{D7} \wedge \cF_{D7}=
3\left(2n_b+1\right)-6\left(1-n_s\right), \nonumber
\ee
since $\cD_{D7'}=\cD_{D7}$ and $\cF_{D7'}=-\cF_{D7}$, while:
\be
I_{D7-O7}=\int_{\cD_{D7}\cap \cD_{O7}}\cF_{D7}=\frac 12\,I_{D7-D7'}\,,
\ee
as $\cD_{O7}=\cD_{D7}$. Hence (\ref{IU4}) reduces to $I_{U(4)}= I_{D7-D7'}$.

\item $\cF_{D7}\neq 0$ induces chiral states between the D7-stack and the E3-instanton.
The number of E3 zero-modes in the fundamental representation of the hidden $U(4)$ gauge group is \cite{Blumenhagen:2006xt,Ibanez:2006da,Florea:2006si, Blumenhagen:2009qh}:
\be
I_{D7-E3}=\int_{\cD_{D7}\cap \cD_s} \left(F_{D7} - F_{\rm E3}\right) = \int_{\cD_{D7}\cap \cD_s} \cF_{D7} =
3\left(1-n_s\right).
\ee
\end{itemize}

The presence of chiral E3 zero-modes can prevent the contribution of the E3 instanton to the superpotential.
In this case we only obtain a contribution if the chiral states develop a non-vanishing VEV \cite{Blumenhagen:2007sm,Blumenhagen:2008zz}. Therefore, to avoid this possible problem, we consider $n_s=1$
so that $I_{D7-E3}$ vanishes. We finally have:
\be
I_{D7-E3}=0\,,\qquad I_{U(4)}=3\left(2n_b+1\right)\qquad\text{and}\qquad \xi_{D7} =\frac{I_{U(4)}}{2}\frac{t_b}{\vo}\,.
\label{FIb}
\ee
The D3-charge of this configuration (including the four O3-planes) is:
\be
Q_{D3}^{\rm hid}= -\frac{n_{O3}}{4}-\frac12 \chi(\cD_{O7}) + 8\left( -\frac12\int_{\cD_{O7}} \cF_{D7}^2 \right)=-19-3(1+2n_b)^2=-22\:,
\label{eq:d3tadpolecond}
\ee
where at the end we set $n_b=-1$.


\section{Moduli stabilisation}
\label{sec:stabilisation}

We have now collected all the ingredients to proceed with the
discussion of moduli stabilisation in the explicit example.
In Section \ref{sec:NumericsEx1} we stabilise of the dilaton and complex
structure moduli using background fluxes.
In Section \ref{sec:KaehlerStab} we come then to the stabilisation of the K\"ahler moduli. Our discussion is based on the geometry
of the CY three-fold $X_3$ introduced in Section~\ref{sec:example}
and the associated brane setup discussed in the previous section. 
The results are compared to the ones obtained for the three-fold $\mathbb{P}^4_{[1,1,1,6,9]}[18]$. 
Equipped with the explicit flux solutions, we then proceed with the stabilisation of the K\"ahler moduli and find explicit de Sitter minima with TeV-scale soft masses.

\subsection{Dilaton and complex structure moduli stabilisation}
\label{sec:NumericsEx1}

In Section~\ref{sec:example} we obtained the pre-potential~\eqref{eq:prepotentialexample} and
all the corresponding periods. Let us now
proceed with the stabilisation of the dilaton and complex structure
moduli.  We turn on flux only along
three-cycles which are  invariant under $\mathbb{Z}_3^3$. 
The complex structure moduli which are not invariant under the
discrete symmetry are
stabilised such that the Calabi-Yau is symmetric under the GP-group $\mathbb{Z}_3^3$,
i.e.~zero, following exactly the arguments outlined in Section
\ref{sec:ReducingCSM}. Hence, they are not present in our prepotential.

We are then left with the stabilisation of the (four) invariant $u$-moduli and the dilaton.
The imaginary self-dual (ISD) condition for the fluxes halves the number of independent flux choices.
We consider in the following fluxes with a positive D3 tadpole, which
is a necessary condition for the ISD condition. We specify fluxes of the form:
\begin{equation}
\begin{pmatrix}
\tilde M_K\\
\tilde N^K
\end{pmatrix}
=\begin{pmatrix}
-N^K\\
M_K
\end{pmatrix},
\end{equation}
which leaves us with $2\times 4+2=10$ independent flux parameters. Indeed, in this parametrisation the D3 tadpole~\eqref{eq:D3tad} becomes automatically positive semi-definite, $Q_{D3}=\sum_K ((\tilde M_K)^2+(\tilde N^K)^2)$. 
Furthermore, some flux configurations are related by $SL(2,\mathbb{Z})$ transformations which leads to further redundancies:
\begin{equation}
\begin{pmatrix}
\tilde M_K\\
\tilde N^K
\end{pmatrix}\simeq
\begin{pmatrix}
-\tilde M_K\\
-\tilde N^K
\end{pmatrix}\simeq
\begin{pmatrix}
-\tilde N^K\\
\tilde M_K
\end{pmatrix}
\simeq
\begin{pmatrix}
\tilde N^K\\
-\tilde M_K
\end{pmatrix}.
\end{equation}
We are interested in $SL(2,\mathbb{Z})$ invariant flux configurations with a tadpole $Q_{D3}\leq 22-2N_{D3}$, since the  D3-tadpole of our setup in Section~\ref{sec:hiddendbranes} was $-(22-2N_{D3})$ (cf.~\eqref{eq:d3tadpolecond} after including the contribution from the D3-branes at the singularities).

Looking at the prepotential~\eqref{eq:prepotentialexample} we observe an additional
symmetry between the complex structure moduli $u_2$ and $u_3$. This is related to the orientifold involution.\footnote{It is worthwhile
to note that the generators $K_2$ and $K_3$,
cf.~\eqref{eq:Kaehlerconeex1}, correspond to the monomials
$\alpha_8\,z_1^3 \, z_2^3 \, z_5^3$ and $\alpha_7\,z_1^3 \, z_3^3 \,
z_6^3$ in the hypersurface equation \eqref{eq:reduced-HSE-model1}.
Therefore, setting $u_2=u_3$ is linked with the involution
\eqref{eq:involution-model1} of the CY three-fold.} To consider orientifold invariant configurations, we need to restrict our analysis to minima with
$u_2=u_3$ and flux choices which have equal fluxes corresponding to
these two moduli. This reduces the problem to eight independent flux
choices and three complex structure moduli.

From the prepotential~\eqref{eq:prepotentialexample} and a given flux choice we can calculate the scalar potential and, in particular, the covariant derivatives $D_i W.$ We are interested in the solutions of~$D_iW=0$ and would like to find such minima numerically as an analytic treatment seems infeasible. To minimise the potential numerically in an efficient and reliable way, we limit ourselves to the prepotential without the instanton corrections at first (in the mirror symmetric language). Given a minimum for this setup, we check in a second step whether the instanton corrections are small enough to keep this minimum stable at the computed values of $u_i$.

To solve this problem numerically, we use the openly available package Paramotopy~\cite{PM}, which allows to scan over various flux choices more efficiently. Paramotopy relies on Bertini~\cite{BHSW06} which is a homotopy continuation solver that produces solutions to polynomial systems. This numerical analysis only identifies isolated minima but neglects continuous ones. As we have to treat the conjugate of a complex structure modulus as an independent variable during the numerical search with this code, we stress that this might neglect various minima which are not found because they might be realised as continuous minima in this larger system under consideration.

To simplify the system further we can solve the dilaton equation explicitly. Indeed, we find that
\beq \label{eq:tauFterm}
	\tau=\frac{f\cdot\bar{\Pi}}{h\cdot\bar{\Pi}}
\eeq
and a similar expression for its complex conjugate $\bar{\tau}$ solve
the F-term condition
\beq
\nonumber 0=\partial_\tau W+W\ \partial_\tau K=-h\cdot\Pi+(f-\tau h)\cdot\Pi \left(-\frac{1}{\tau-\bar{\tau}}\right)\,,
\eeq
where we used the K\"ahler potential \eqref{eq:Kcs+tau} and the
notation introduced in \eqref{eq:KaehlerpotCS}.

We can use this result to simplify the covariant derivatives for the
complex structure moduli as follows. First we expand
the first equation in \eqref{eq:F-term}  explicitly as
\beq
\nonumber 0=\partial_{i} W+W\ \partial_{i} K= (f-\tau h)\cdot\partial_{i}\Pi-(f-\tau h)\cdot\Pi\ \frac{\Pi^\dagger\cdot\Sigma\cdot\partial_{i}\Pi}{\Pi^\dagger\cdot\Sigma\cdot\Pi}\,.
\eeq
Next, we cancel the denominator, then use \eqref{eq:tauFterm} and
cancel denominators again to obtain
\beq
	0=(\Pi^\dagger\cdot\Sigma\cdot\Pi)\left(f\cdot\partial_i\Pi\ h\cdot\bar{\Pi}-f\cdot\bar{\Pi}\ h\cdot\partial_{i}\Pi\right)-\left(f\cdot\Pi\ h\cdot\bar{\Pi}-f\cdot\bar{\Pi}\ h\cdot\Pi\right)\ (\Pi^\dagger\cdot\Sigma\cdot\partial_{i}\Pi)\,.
\nonumber
\eeq
We use this equation as an input for the numerical minimisation which
is a set of polynomial equations. In complete analogy we can obtain the
input for the covariant derivatives with respect to the complex
conjugated variables.

\begin{figure}[ht!]\begin{center}
 \includegraphics[width=0.7\textwidth]{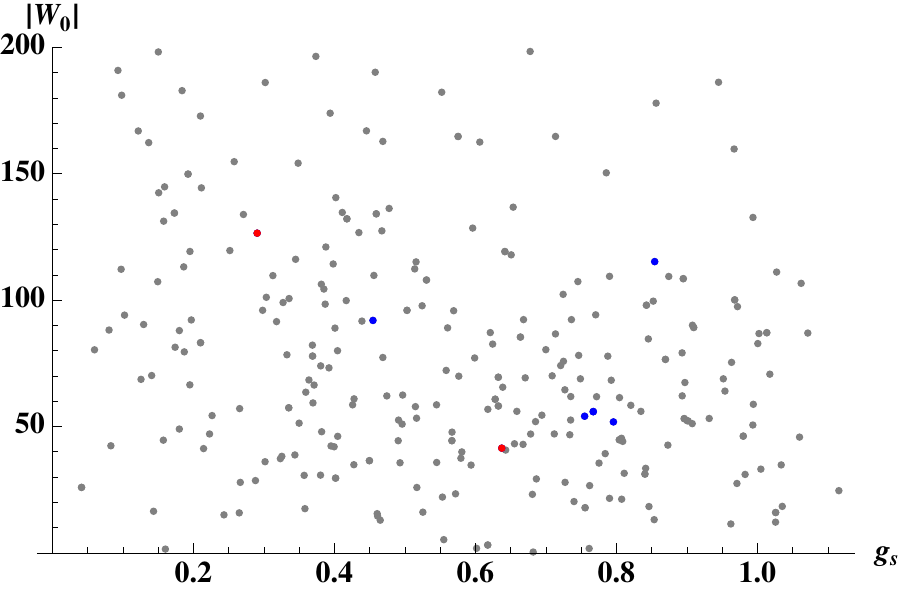}
\captionof{figure}{\footnotesize{Distribution of flux solutions in the $(g_s,|W_0|)$-plane in the fundamental domain after performing appropriate $SL(2,\mathbb{Z})$ transformations.}\label{fig:dist2}}
\end{center}\end{figure}

Under the above assumptions, we have searched for minima for each D3 tadpole in the range between $Q_{D3}=10,\ldots,20$ for 100 randomly
chosen flux configurations. On average we find $2.59$ solutions for a single flux configuration.
At first many solutions correspond to solutions outside the fundamental domain and with large instanton corrections, but we can use the $SL(2,\mathbb{Z})$ freedom to transform solutions into the fundamental domain. We show in Figure~\ref{fig:dist2} the
distribution in the $(g_s,|W_0|)$-plane after restricting to the fundamental domain: Most of the points (shown in gray) correspond to points with large instanton contributions, blue and red points have small instanton contributions (satisfying $|F_{\rm inst}|/|F|<0.1$ and ${\rm max}_i\left(|F^i_{\rm inst}|\right)/|F|<0.1$) taking instanton corrections up-to order 2 (blue) and order 10 (red) into account.

\begin{figure}[ht!]\begin{center}
 \includegraphics[width=0.48\textwidth]{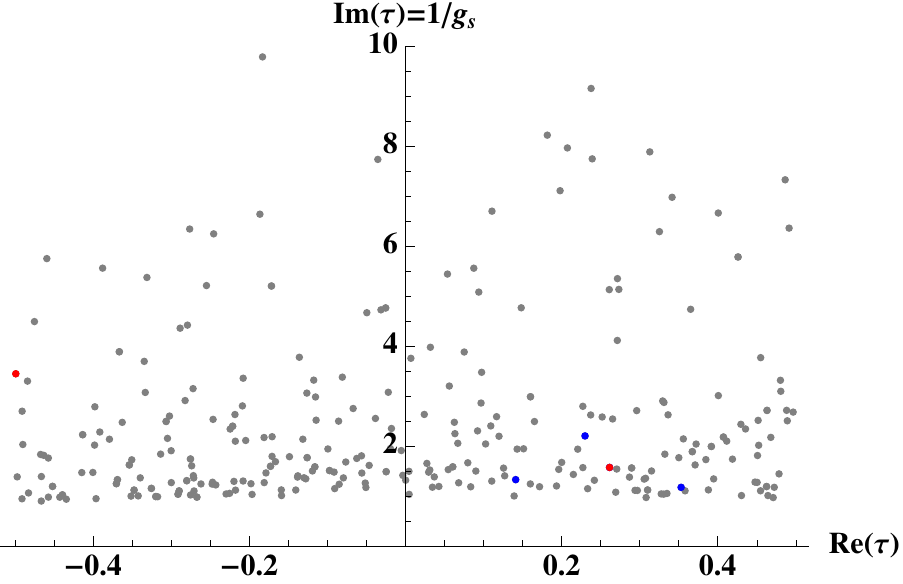}\hspace{0.2cm}
  \includegraphics[width=0.48\textwidth]{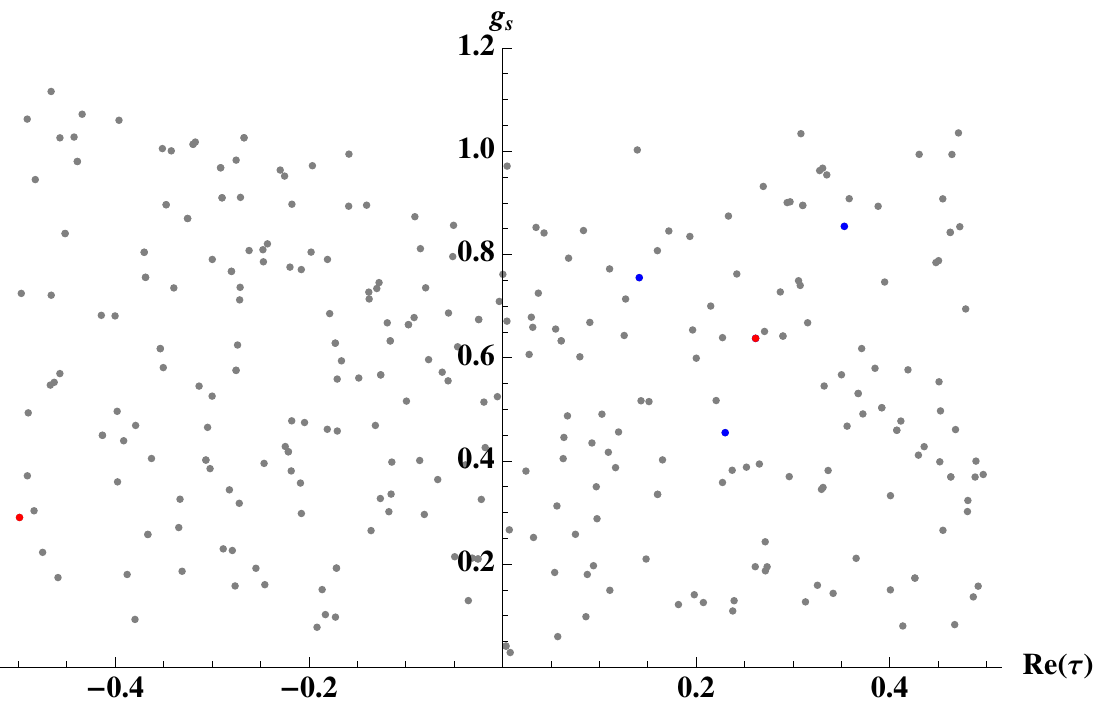}\\
   \includegraphics[width=0.5\textwidth]{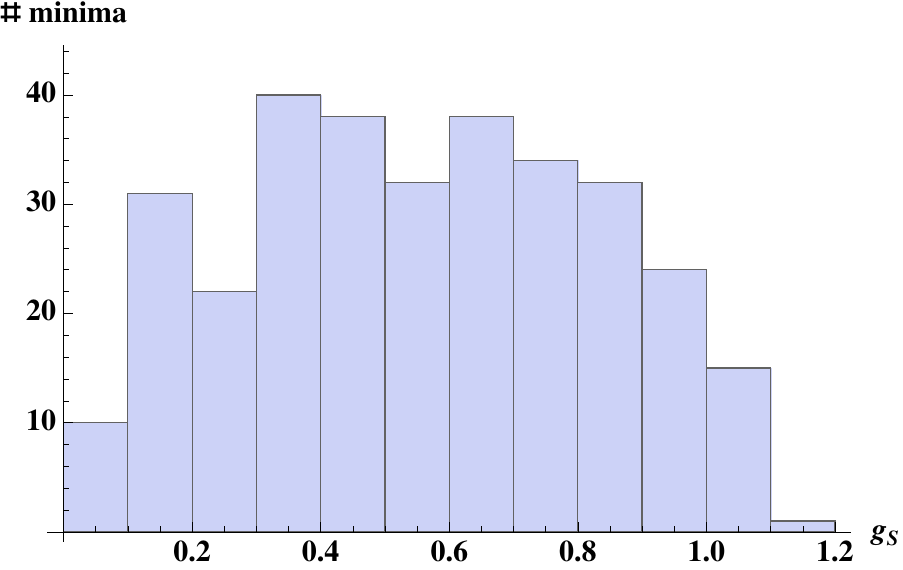}
\captionof{figure}{\footnotesize{Distribution of solutions in the fundamental domain of $\tau,$ i.e.~in the $(\text{Re}[\tau],\text{Im}[\tau]=1/g_s)$ plane (top left). Next to it (top right) we show the distribution in the $(\text{Re}[\tau],1/\text{Im}[\tau]=g_s)$ plane. The bottom shows the distribution of minima with respect to values of $g_s$ before restricting to small instanton corrections. Gray points correspond to points with large instanton corrections, blue and red points correspond to small instanton contributions taking corrections up to order 2 and 10 into account. \label{fig:distgs}}}
\end{center}\end{figure}

The distribution of minima
(again before invoking small instanton contributions) in the fundamental domain is shown in Figure~\ref{fig:distgs}.

Finally, in Table~\ref{tab:physmin} and~\ref{tab:physmin2}, we
display some minima with small instanton contributions. As explained
above this is a necessary requirement for retaining the minimum
obtained from the classical terms also in the presence of the instanton
corrections.
\begin{table}
\begin{footnotesize}
\begin{tabular}{c|c|c|c | c |c| c| c}
$Q_{D3}$ & $(\tilde N_i,\tilde M_i)$ & $u_1$ & $u_{2,3}$ & $u_4$ & $\tau$ & $g_s$ & $|W_0|$\\ \hline\hline
19 & (1,-3,0,0,0,0,2,-1) & $4.39-1.22 i$ & $19.3+0.971 i$ & $-21.6+1.18 i$ & $2.21+2.58 i$ & $0.39$ & $954.4$\\\hline
14 & (1,1,1,1,3,0,0,0) & $-5.99+2.69 i$ & $4.09+1.59 i$ & $-0.115-0.0581 i$ & $-2.76+1.24 i$ & $0.8$ & $82.66$\\\hline
14 & (1,0,0,1,1,-3,1,0) & $4.72+2.7 i$ & $-3.92+1.94 i$ & $0.176-0.0468 i$ & $4.14+1.32 i$ & $0.75$ & $54.03$\\\hline
15 & (2,0,2,1,0,0,-1,0) & $28.+3.3 i$ & $-11.4+2.62 i$ & $0.331-0.0291 i$ & $6.72+1.3 i$ & $0.77$ & $55.85$\\\hline
15 & (1,2,1,1,1,2,-1,0) & $1.49+0.861 i$ & $-1.22+1.77 i$ & $-0.201-0.0276 i$ & $-2.41+2.22 i$ & $0.45$ & $36.44$\\\hline
18 & (1,2,0,2,2,2,0,-1) & $1.13+0.473 i$ & $-0.327+2.02 i$ & $-0.583+0.103 i$ & $-1.5+3.44 i$ & $0.29$ & $126.5$
\end{tabular}
\caption{\footnotesize Summary of all solutions which satisfy $0<g_s<1$ and obey $|F_{\rm inst}|/|F|<1,$ ${\rm max}_i\left(|F^i_{\rm inst}|\right)/|F|<1.$ The flux configuration is given where the third and respectively seventh entry denote the flux quanta $\tilde N_2=\tilde N_3$ respectively $\tilde M_2=\tilde X_3$ which are chosen to be the same as described in the main text. \label{tab:physmin}}
\end{footnotesize}
\end{table}

\begin{table}
\begin{footnotesize}
\begin{tabular}{c|c|c|c|c|c|c}
$Q_{D3}$ & $(\tilde N_i,\tilde M_i)$ & $g_s$ & $W_0$ & $|W_0|$ &$|F_{\rm inst}|/|F|$& ${\rm max}_i\left(|F^i_{\rm inst}|\right)/|F|$\\ \hline\hline
19 & (1,-3,0,0,0,0,2,-1) & $0.39$ & $618.+727.3 i$ & $954.4$ & $0.0744$ & $0.00976$\\\hline
14 & (1,1,1,1,3,0,0,0) & $0.8$ & $-41.33-71.58 i$ & $82.66$ & $0.731$ & $0.0592$\\\hline
14 & (1,0,0,1,1,-3,1,0) & $0.75$ & $-9.465+53.2 i$ & $54.03$ & $0.723$ & $0.051$\\\hline
15 & (2,0,2,1,0,0,-1,0) & $0.77$ & $49.89+25.11 i$ & $55.85$ & $0.0114$ & $0.00387$\\\hline
15 & (1,2,1,1,1,2,-1,0) & $0.45$ & $-30.06-20.59 i$ & $36.44$ & $0.788$ & $0.0968$\\\hline
18 & (1,2,0,2,2,2,0,-1) & $0.29$ & $-120.3-38.86 i$ & $126.5$ & $0.014$ & $0.0257$
\end{tabular}
\caption{\footnotesize Examples of solutions which satisfy $0<g_s<1$ and obey $|F_{\rm inst}|/|F|<1,$ ${\rm max}_i\left(|F^i_{\rm inst}|\right)/|F|<0.1.$ Here we show the respective size of their instanton contributions. Again note that the third and respectively seventh entry for the flux configuration denote the entries for $\tilde N_2=\tilde N_3$ respectively $\tilde M_2=\tilde M_3$ which are chosen to be the same as described in the main text. $F^i_{\rm inst}$ denotes the leading order contribution in $F_{\rm inst}$ in a given $u_i$.\label{tab:physmin2}}
\end{footnotesize}
\end{table}

In order to have a consistent superspace derivative expansion in the 4D effective field theory, one requires $\vo \gg |W_0|^3$ \cite{Cicoli:2013swa}.
This implies that the K\"ahler moduli have to be stabilised such that $\vo \gg 10^{3}$, respectively, $\vo \gg 10^{6}$ depending on $|W_0|$.

\begin{figure}[ht!]\begin{center}
 \includegraphics[width=0.48\textwidth]{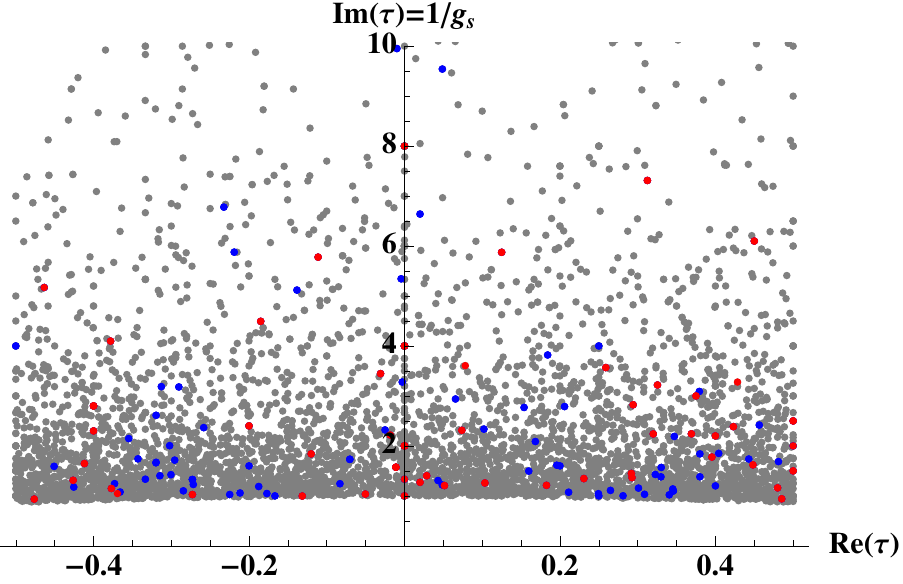}\hspace{0.2cm}
  \includegraphics[width=0.48\textwidth]{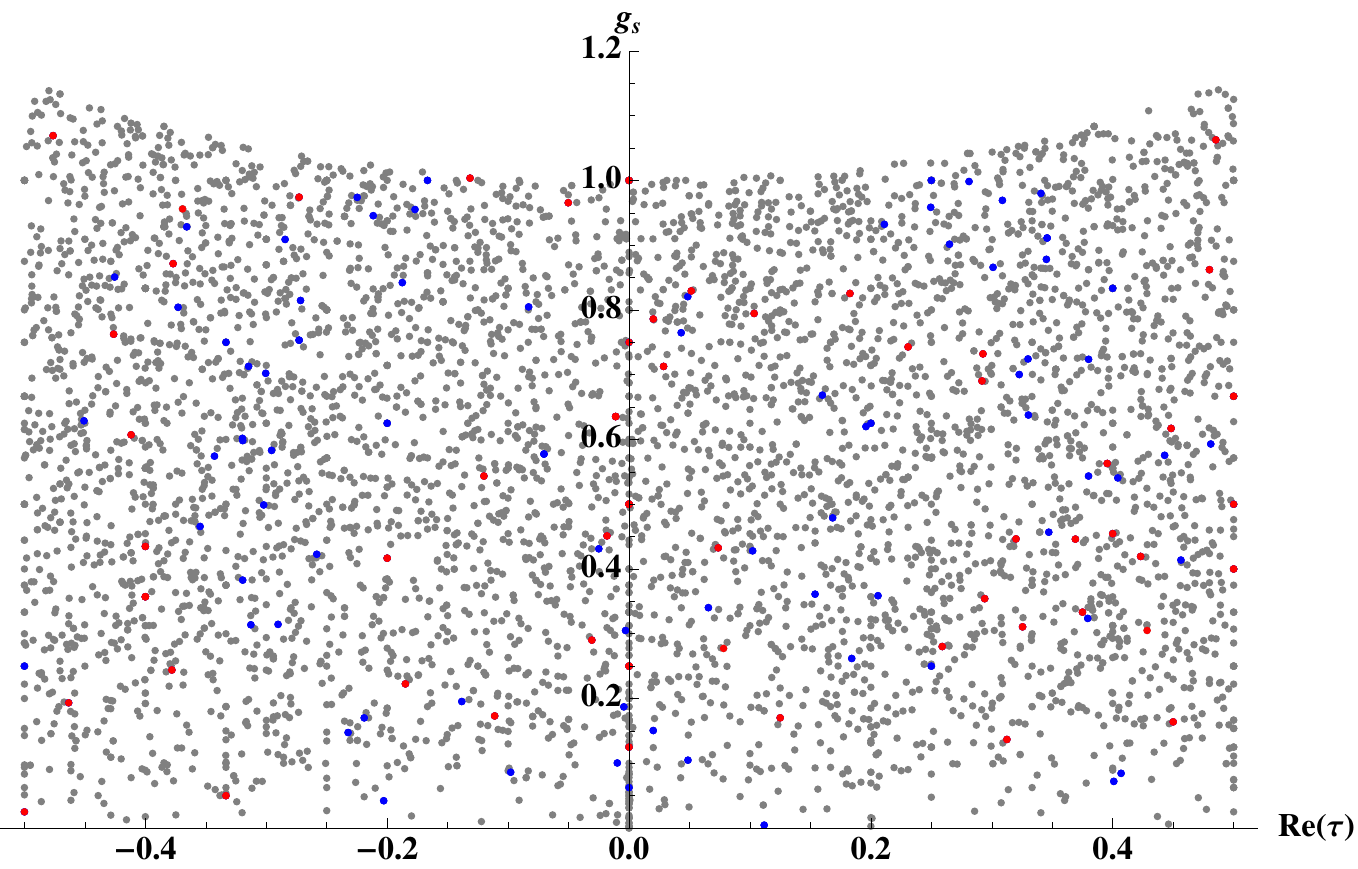}\\
   \includegraphics[width=0.5\textwidth]{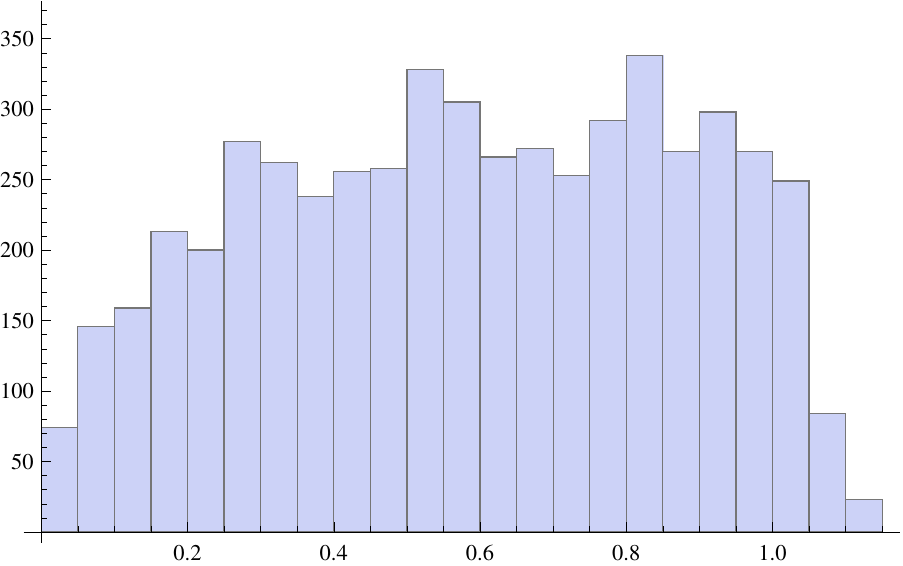}
\captionof{figure}{\footnotesize{The same distributions as before but for $\mathbb{P}_{1,1,1,6,9}.$ Distribution of solutions in the fundamental domain of $\tau,$ i.e.~in the $(\text{Re}[\tau],\text{Im}[\tau]=1/g_s)$ plane (top). In the middle we show the distribution in the $(\text{Re}[\tau],1/\text{Im}[\tau]=g_s)$ plane. The bottom shows the distribution of minima with respect to values of $g_S$ before restricting to small instanton corrections. Gray points correspond to points with large instanton corrections, blue and red points correspond to small instanton contributions taking corrections up to order 2 and 14 into account (again requiring $|F_{\rm inst}|/|F|<0.1$ and ${\rm max}_i\left(|F^i_{\rm inst}|\right)/|F|<0.1$). \label{fig:dist5}}}
\end{center}\end{figure}

To compare our results to previous analysis, we show in Figure~\ref{fig:dist5} 
the corresponding distributions for $\mathbb{P}^4_{[1,1,1,6,9]}[18]$. Note that 
the gray points in Figures~\ref{fig:dist2} and~\ref{fig:distgs} are not 
solutions of $DW=0$ when $W$ is the full type IIB superpotential. On the other 
hand, they satisfy $D\hat{W}=0$, where $\hat{W}$ is the cubic polynomial  in 
the $u_i$ arising from the prepotential without instanton corrections as 
in~\eqref{eq:prepotentialexample}. We can consider $\hat{W}$ as the 
superpotential for a toy model. From the Figures~\ref{fig:dist2} 
and~\ref{fig:distgs}, we see that in this toy model we find results consistent 
with a uniform distribution of $W_0$ and $g_s.$

\subsection{K\"ahler moduli stabilisation}
\label{sec:KaehlerStab}

As we have seen in the previous section, the dilaton and the complex
structure moduli can be fixed supersymmetrically at semi-classical level
by turning on quantised background fluxes.
For concreteness, we shall focus on the flux vacuum related to the last
line in Table~\ref{tab:physmin}: the string coupling is in the
perturbative regime, $g_s=0.29$, the VEV of the flux superpotential is
$|W_0|=126.5$ and the D3-charge of the flux is $Q_{D3}=18$, leaving
space for $N_{D3}\leq2$ D3-branes on top of the dP$_6$ singularities.

\subsubsection*{Definition of the K\"ahler moduli}

Due to the no-scale property of the type IIB K\"ahler potential for the
K\"ahler moduli $T_i$, $i=1,...,h^{1,1}(X_3)$,
$K_{T-{\rm moduli}} = - 2 \ln\vo$, the fluxes do not induce any
tree-level F-term potential for the $T$-moduli.
The K\"ahler moduli can however be lifted by either perturbative or
non-perturbative effects. A general strategy to lift these remaining
flat directions in a way compatible with the presence of chirality
has been described in~\cite{Cicoli:2011qg,Cicoli:2012vw}. Here we shall
simply apply the same stabilisation scheme to our particular example.

The total number of K\"ahler moduli is $h^{1,1}(X_3)=4$. Three
of them are even under the orientifold and are defined as:
\be
T_b=\tau_b+{\rm i}\left(\int_{\cD_b}C_4\right), \qquad T_s=\tau_s+{\rm i}\left(\int_{\cD_s}C_4\right),
\qquad T_{\rm vs}=\tau_{\rm vs}+{\rm i}\left(\int_{\cD_{q_1} + \cD_{q_2}}C_4\right), \nn
\ee
where the visible sector four-cycle volume is
$\tau_{\rm vs}=\tau_{q_1} + \tau_{q_2}$.
On the other hand, the orientifold odd K\"ahler modulus is defined as
\cite{Grimm:2004uq}:
\be
G=b + {\rm i}c \equiv \left(\int_{\cD_{q_1} - \cD_{q_2}}B_2\right)
+{\rm i}\left(\int_{\cD_{q_1} - \cD_{q_2}}C_2\right).
\ee

\subsubsection*{Leading order stabilisation}

The potential for the K\"ahler moduli can be organised in an expansion
in inverse powers of the overall CY volume $\vo$.
As we shall see, the leading order contribution comes from D-terms
which depend on the K\"ahler moduli
because gauge fluxes induce $T$-dependent FI-terms.

In addition to the anomalous $U(1)$ living on the stack of D7-branes on top of the O7-plane, there are two additional anomalous $U(1)$
symmetries belonging to the gauge theory located at the
dP$_6$ singularity.\footnote{The gauge theory associated to any dP$_n$
singularity, with $n=0,1,...,8$, gives always rise to two anomalous
$U(1)$'s.} Thus the total D-term potential looks like:
\be
V_D = V_D^{\rm quiver} + V_D^{\rm bulk}\,,
\ee
where:
\be
V_D^{\rm quiver} = \frac{1}{{\rm Re}(f_{D3_1})} \left(\sum_i q_{D3_1 i} |C_i|^2 -\xi_{D3_1} \right)^2
+ \frac{1}{{\rm Re}(f_{D3_2})} \left( \sum_i q_{D3_2 i} |C_i|^2 -\xi_{D3_2}\right)^2,
\label{Dpot}
\ee
and:
\be
V_D^{\rm bulk} = \frac{1}{{\rm Re}(f_{D7})} \left(\sum_i q_{D7i} |\phi_i|^2 - \xi_{D7}\right)^2.
\label{Dpotbulk}
\ee
In the previous expressions, the gauge kinetic functions are given by:
\be
f_{D3_1}= \tau + q_{D3_1} T_{\rm vs}\,,\qquad f_{D3_2}= \tau + q_{D3_2} G \qquad\text{and}\qquad f_{D7} = T_b -T_s + k \tau\,,
\ee
where $q_{D3_i}$ are the $U(1)$ charges of $T_{\rm vs}$ and $G$ while
$k$ is a parameter which depends on $\cF_{D7}$.
The $C_i$ and $\phi_i$ are canonically normalised matter fields with
$U(1)$ charges $q_{D3_1 i}$, $q_{D3_2 i}$ and $q_{D7i}$
living respectively on the fractional D3- and D7-branes.\footnote{Note that we denoted
the fields $X$, $Y$, $Z$ of Section 3.2 as $C_i$, while the $\phi_i$ are fields in the
anti-symmetric representation of $U(4)$, cf (56).}
The FI-terms $\xi_{D3_1}$ and $\xi_{D3_2}$ read
\cite{Grana:2003ek}:
\be
\xi_{D3_1} = 4 q_{D3_1}\,\frac{\tau_{\rm vs}}{\vo}\qquad\text{and}\qquad \xi_{D3_2} = 4 q_{D3_2}\,\frac{b}{\vo}\,,
\ee
while $\xi_{D7}$ is given in (\ref{FIb}).

One can easily see that the D-term potential scales as
$V_D^{\rm quiver}\sim V_D^{\rm bulk} \sim \mc{O}(1/\vo^2)$.
This is of the same order as the flux-generated F-term potential used
to fix $\tau$ and all the complex structure moduli:
\be
V_F^{\rm flux} \simeq \frac{1}{\vo^2}\left(|D_\tau W|^2+|D_u W|^2\right).
\ee
In order to have a consistent stabilisation procedure, we have
therefore to set $V_D=0$ so that
supersymmetry is preserved at this order of expansion in inverse powers
of $\vo$. This condition gives:
\be
\xi_{D3_1}=\sum_i q_{D3_1i} |C_i|^2\,,\qquad \xi_{D3_2}=\sum_i q_{D3_2 i} |C_i|^2\qquad\text{and}\qquad
\xi_{D7}=\sum_i q_{D7 i} |\phi_i|^2\,.
\label{Relations}
\ee
These relations fix only three combinations of closed and open string
moduli, corresponding to
the combinations of axions which get eaten up by the anomalous $U(1)$s
via the St\"uckelberg mechanism.
The mass of each Abelian gauge boson is proportional to the open and
closed string axion decay constants $f_{{\rm ax},j}^{\rm op}$ and $f_{{\rm ax},j}^{\rm cl}$ \cite{ArkaniHamed:1998nu}:
\be
M_{U(1),j}^2 \simeq \frac{1}{{\rm Re}(f_j)} \left[\left(f_{{\rm ax},j}^{\rm op}\right)^2+\left(f_{{\rm ax},j}^{\rm cl}\right)^2\right]\,,
\label{MU(1)}
\ee
where:
\be
\left(f_{{\rm ax},j}^{\rm op}\right)^2 \simeq \xi_j
\qquad\text{and}\qquad \left(f_{{\rm ax},j}^{\rm cl}\right)^2 \simeq \frac{\partial^2 K}{\partial T_j^2}\,.
\label{fs}
\ee
For the anomalous $U(1)$ on the D7-stack in the bulk, we have:
\be
\left(f_{{\rm ax},D7}^{\rm op}\right)^2 \simeq \xi_{D7} \simeq \frac{1}{\tau_b}\gg
\left(f_{{\rm ax},D7}^{\rm cl}\right)^2 \simeq \frac{\partial^2 K}{\partial T_b^2} \simeq \frac{1}{\tau_b^2}
\qquad\text{for}\qquad\tau_b\sim\vo^{2/3}\gg 1\,,
\ee
implying that the combination of moduli fixed by D-terms is mostly
given by open string modes.
We can therefore consider $\tau_b$, or equivalently $\vo$, as a flat
direction at this level of approximation.
As can be seen from (\ref{MU(1)}), the anomalous $U(1)$ develops a mass of the order of the Kaluza-Klein
scale: $M_{U(1),D7}\sim M_P/\tau_b \sim M_P/\vo^{2/3}\sim M_{KK}$.

On the other hand, the Abelian gauge bosons at the dP$_6$ singularity
become massive by eating up the closed string axions
since:
\be
K\supset \frac{\left(T_{\rm vs}+\bar{T}_{\rm vs}\right)^2}{\vo}\quad\Rightarrow\quad\left(f_{{\rm ax},D3_1}^{\rm op}\right)^2
\simeq \frac{\tau_{\rm vs}}{\vo}\ll
\left(f_{\rm ax,D3_1}^{\rm cl}\right)^2 \simeq \frac{\partial^2 K}{\partial T_{\rm vs}^2}\simeq \frac{1}{\vo}
\quad\text{for}\quad\tau_{\rm vs}\ll 1\,, \nn
\ee
and similarly for the second $U(1)$. Hence the D-terms from the quiver
construction fix $\tau_{\rm vs}$ and $b$.
The corresponding axions get eaten up by the two anomalous $U(1)$s
which from (\ref{MU(1)}) acquire a mass of the order of the string scale:
$M_{U(1),D3} \sim M_P/\vo^{1/2}\sim M_s$. Similarly to $\tau_b$, the
matter fields $|C_i|$ remain massless at this level of approximation.

\subsubsection*{Subleading order stabilisation}

In order to fix $T_b$, $T_s$ and $|C_i|$ we have to consider subleading
F-term contributions.

\medskip
\emph{Visible sector matter fields}
\medskip

Focusing on the stabilisation of the visible sector matter fields, we
have (we restrict to the case of just one open string field $C$ where
we set the VEV of its phase to zero without loss of generality\footnote{A different value of the phase could change the sign of $c_3$, for example, but
given that we are not working out the exact sign of $c_3$, the exact value of the phase is irrelevant for our conclusions.}):
\bea
V_F (|C|) &=&  c_2 m_0^2 |C|^2 + c_3 A |C|^3 + c_4 \lambda |C|^4 + \mc{O}(|C|^5) - d_4\,\frac{\tau_{\rm vs}^2}{\vo^3}\left[1+\mc{O}\left(\frac{1}{\vo}\right)\right] \nonumber \\
&\simeq& c_2 \frac{|C|^2}{\vo^{\alpha_2}}  + c_3 \frac{|C|^3}{\vo^{\alpha_3}}
+ \left(\frac{c_4}{\vo^{\alpha_4}} - \frac{d_4}{\vo} \right) |C|^4 \,,
\label{VFC}
\eea
where the first three terms come from expanding the F-term scalar
potential in powers of $|C|$ up to
fourth order, while the last term
comes from the breaking of the no-scale structure by $\tau_{\rm vs}$
which we wrote in terms of $|C|$ using the first relation in
(\ref{Relations}). In more detail, the quadratic term is proportional
to the soft scalar mass of $|C|$
which we have parametrised as $m_0^2\simeq \vo^{-\alpha_2}$, the cubic
term is proportional to the soft $A$-term
parametrised as $A \simeq \vo^{-\alpha_3}$, while the quartic term is
proportional to the coupling $\lambda \simeq \vo^{-\alpha_4}$.

Depending on the sign of $c_2$, the VEV of $|C|$ can be zero or
non-vanishing. If $c_2>0$, then $|C|=0$, implying, by
\eqref{Relations}, $\xi_{D3,i}=0$ $\forall i=1,2$ which fixes
$\tau_{\rm vs} = b= 0$. This ensures that the dP$_6$ divisor supporting
the visible sector is really collapsed to zero size.

If instead the matter field $|C|$ develops a tachyonic mass from
supersymmetry breaking, i.e. $c_2<0$,
then, depending on the signs of the different coefficients in
(\ref{VFC}), $|C|$ can develop a non-zero VEV.
In this case, one may worry that $\tau_{\rm vs}$ gets fixed at values
larger than the string scale, so
resolving the dP$_6$ singularity. However this is not the case. In
fact, for models with just fractional D3-branes,
the visible sector is sequestered from supersymmetry breaking,
leading to $\alpha_3=2$, $\alpha_4=1$ and either $\alpha_2=3$ or
$\alpha_2=4$, depending on the moduli-dependence of the K\"ahler metric
for matter fields~\cite{Blumenhagen:2009gk}.
Thus $|C| \simeq M_P /\vo^{\alpha_2-2}$, implying from
(\ref{Relations}) that $\tau_{\rm vs}\simeq \vo^{5-2\alpha_2}$, with a
similar relation holding for $b$.
We then see that for $\alpha_2=3$ $\tau_{\rm vs} \sim \vo^{-1}\ll 1$,
while for $\alpha_2=4$ $\tau_{\rm vs}\sim \vo^{-3}\ll 1$,
and so in both cases the dP$_6$ divisor is still in the singular
regime.

An interesting implication of this case is that the phase $\theta$ of
$C=|C|\,e^{{\rm i}\theta}$ can behave as the QCD axion for
$\alpha=3$. In fact, its decay constant $f_{\rm ax}^{\rm op}$ is set by
the above VEV of $|C|$ which can take two different values
(using $\vo \sim 10^8$ as derived below):
\be
f_{\rm ax}^{\rm op} \simeq \frac{M_P}{\vo}\simeq 10^{10}\,\text{GeV} \quad\text{for}\quad \alpha_2=3
\qquad\text{and}\qquad
f_{\rm ax}^{\rm op} \simeq \frac{M_P}{\vo^2}\simeq 100\,\text{GeV}\quad\text{for}\quad \alpha_2=4\,. \nn
\ee
The value of $f_{\rm ax}^{\rm op}$ for $\alpha_2=3$ lies inside the
phenomenologically allowed window for the QCD axion:
$10^9\,\text{GeV}\lesssim f_{\rm QCD}\lesssim10^{12}\,\text{GeV}$ while
the case with $\alpha_2=4$ seems to be ruled out.
Note that $C$ has to be a SM singlet due to the large value of its VEV.
Moreover, there are two anomalous $U(1)$s at the dP$_6$ singularity,
and so only one combination of open string axions can play the r\^ole
of the QCD axion and get massive by QCD instantons
whereas the other combination would remain massless.

In the presence of flavour D7-branes, threshold effects de-sequester
the visible sector inducing soft terms of order
the gravitino mass, and so $\alpha_2=2$. In this case $c_2$ has to be
positive otherwise $\tau_{\rm vs}$ would develop
a VEV of order unity, thus resolving the dP$_6$ singularity. The
matter field $|C|$ would therefore have a vanishing VEV
which could however become non-zero via RG running below the string
scale. $\tau_{\rm vs}$ would then develop a very small VEV compatible
with the singular regime and one could obtain a viable QCD axion
from the phase of a matter field analogous to the situation described
above. A similar mechanism could help to have a viable QCD axion
also in the case with just fractional D3-branes and $\alpha_2=4$.

\medskip
\emph{K\"ahler moduli in the geometric regime}
\medskip

The scalar potential for the K\"ahler moduli $T_s$ and $T_b$ receives
three different leading order contributions:
$V_{\rm np}$ from non-perturbative superpotential effects,
$V_{\alpha'}$ from perturbative corrections to the K\"ahler potential
and soft terms $V_{\rm soft}$  from matter fields living on bulk D7-branes.

The expression for $V_{\rm np}$ is \cite{Balasubramanian:2005zx}:
\be
V_{\rm np} = \frac{8}{3\lambda}  (a_s A_s)^2  \sqrt{\tau_s}\, \frac{e^{-2\, a_s \tau_s}}{\vo}
   - 4 \,a_s A_s |W_0| \tau_s\frac{e^{-a_s \tau_s}}{\vo^2}\,,
\ee
where we have already minimised with respect to the $T_s$-axion and we
have traded $\tau_b$ for $\vo\simeq \lambda\tau_b^{3/2}$.
This potential is generated by an E3-instanton wrapping
$\cD_s$ with $a_s=2\pi$, while the prefactor $A_s$ is an unknown function
of the stabilised complex structure moduli. This instanton contributes
to the superpotential if the $B$-field is chosen as in
(\ref{Bchoice2}). If one instead performs the flux choice
(\ref{Bchoice1}), gaugino condensation in the $SO(8)$ pure super
Yang-Mills theory on the bulk D7-branes gives rise to a tiny
contribution proportional to $e^{-\frac{\pi}{3} \tau_b}$ which can be
neglected. Moreover, a possible contribution from a
rank-two instanton on $\cD_s$ would also be negligible for $\tau_s>1$.

The contribution $V_{\alpha'}$ takes the form \cite{Becker:2002nn}:
\be
V_{\alpha'} = \frac{3}{4} \frac{\zeta |W_0|^2}{g_s^{3/2} \vo^3}\,,
\ee
where $\zeta = -\chi(X_3)\zeta(3) /[2 (2\pi)^3]\simeq 0.32$
using \eqref{eq:HodgeEuler} for the Euler number of $X_3$.
Further perturbative corrections arise due to string loop effects
\cite{Berg:2005ja,vonGersdorff:2005bf}.\footnote{For recent activity on
stringy perturbative corrections
see~\cite{Anguelova:2010ed,Grimm:2012rg,GarciaEtxebarria:2012zm,Grimm:2013gma,Halverson:2013qca}.}
Because of the `extended no-scale structure'
\cite{Berg:2007wt,Cicoli:2007xp}, these corrections
are negligible since they scale as $\vo^{-10/3}$ whereas $V_{\alpha'}$
behaves as $\vo^{-3}$.

The last contribution $V_{\rm soft}$ reads:
\be
V_{\rm soft} = c \,\frac{|W_0|^2}{\vo^2\left[\ln\left(\vo/|W_0|\right)\right]^2} |\phi|^2\,,
\ee
where the soft scalar mass for the matter field $\phi$ living in the
bulk scales as the gravitino mass
$m_{3/2} = e^{K/2} |W| \simeq |W_0| /\vo$ suppressed by a factor of
order $\ln(M_P/m_{3/2})$~\cite{Conlon:2006us,Conlon:2006wz}
($c$ is an $\mc{O}(1)$ coefficient).
Employing the last relation in (\ref{Relations}) one obtains:
\be
V_{\rm soft} = p\,\frac{|W_0|^2}{\vo^{8/3}\left[\ln\left(\vo/|W_0|\right)\right]^2}\,,
\quad\text{with}\quad p = \frac{c\, I_{U(4)}}{2^{1/3} q_{D7}}\,,
\label{Vsoft}
\ee
with $I_{U(4)}$ given in \eqref{IU4}.

Minimising the total F-term potential $V_F^{\rm tot} = V_{\rm np} + V_{\alpha'} + V_{\rm soft}$ with
respect to $\tau_s$ in the limit $\epsilon \equiv 1/(a_s\tau_s)\ll 1$, we obtain:
\be
e^{-a_s \tau_s}\simeq \frac{3\lambda\sqrt{\tau_s}}{4 a_s A_s}\frac{|W_0|}{\vo}
\qquad\Rightarrow\qquad \tau_s (\vo) \simeq\frac{1}{a_s}\ln\left(\frac{\vo}{|W_0|}\right)\,.
\label{tsVEV}
\ee
Plugging this result into $V_F^{\rm tot}$, we find:
\be
V \simeq \frac{|W_0|^2}{\vo^3} \left[\frac{3\zeta}{4 g_s^{3/2}}
-\frac{3\lambda}{2}\tau_s (\vo)^{3/2}
   + p\,\frac{\vo^{1/3}}{\left[a_s \tau_s (\vo)\right]^2} \right].
\label{VFfinal}
\ee
The minimisation with respect to $\vo$ then gives:
\be
\frac{3\zeta }{4g_s^{3/2}}\simeq\frac{3\lambda}{2} \tau_s (\vo)^{3/2}
\left(1-\frac{1}{2a_s \tau_s (\vo)}\right)-\frac 89
\,p\,\frac{\vo^{1/3}}{\left[a_s \tau_s (\vo)\right]^2}\,.
\label{VolVEV}
\ee
Substituting this result in (\ref{VFfinal}), the leading order contribution to the vacuum energy becomes:
\be
\langle V_F^{\rm tot} \rangle \simeq \frac{|W_0|^2}{\langle\vo\rangle^3} \sqrt{a_s \tau_s (\vo)}\left[-\frac{3\lambda}{4\,a_s^{3/2}}
+ \frac{p}{9}\frac{\langle\vo\rangle^{1/3}}{\left[a_s \tau_s (\vo)\right]^{5/2}}\right].
\label{Vmin}
\ee
Setting this expression equal to zero and plugging the result for $p$ back in (\ref{VolVEV}), we find:
\be
\frac{3\zeta }{4g_s^{3/2}}\simeq\frac{3\lambda}{2} \tau_s^{3/2}
\left(1 + \mc{O}(\epsilon)\right)\,.
\label{VolVEV2}
\ee
For $g_s=0.29$,
$|W_0|=126.5$ and $\lambda = \frac 13\sqrt{\frac{2}{3}}$ from
(\ref{vol}), (\ref{VolVEV2})
yields $\tau_s \simeq 2.42$ which fixes the divisor $\cD_s$ supporting the
E3-instanton  in the geometric
regime above the string scale. Setting $a_s=2\pi$, the relation
(\ref{tsVEV}) then yields $\vo \simeq 2.55\cdot 10^7/A_s$.

We can fix the unknown coefficient $A_s$
by imposing TeV-scale supersymmetry. If there
are no flavour branes and the visible sector is sequestered,
the gaugino masses scale as \cite{Blumenhagen:2009gk}:
\be
M_{1/2} = \frac{3\gamma\zeta\,e^{K_{\rm cs}/2}}{4\sqrt{2}g_s}\frac{|W_0| M_P}{\vo^2}\simeq 36.5\,\text{TeV}\,\gamma\,A_s^2,
\ee
where $\gamma$ is an $\mc{O}(1)$ function of the complex structure
moduli which determines the shift of the dilaton VEV due to $\alpha'$
effects, and $e^{K_{\rm cs}}\simeq 1/55.9$ by direct computation in our explicit example.
If $\gamma=1$ and $A_s=0.2$, one obtains $M_{1/2}\simeq 1.5$
TeV. The CY volume is then $\vo\simeq 1.3\cdot 10^8$ while
the parameter $p$ becomes $p\simeq 0.2$. As can be seen from
(\ref{Vsoft}), $p$ depends just on the flux coefficient $n_b$
introduced in \eqref{eq:fluxD7},
the $U(1)$ charge $q_{D7}$ and the parameter $c$, and so we do not have
enough freedom to tune the cosmological constant.
This was expected from the beginning since we chose a particular value
of the flux superpotential $|W_0|$.
However, $p$ turns out to be of order unity, implying that in order to
find a solution with vanishing cosmological constant
one has to vary $|W_0|$ around the value considered in this example,
$|W_0|=126.5$.

If the visible sector is not sequestered from supersymmetry breaking,
then the soft terms become of the order of the gravitino mass and
a larger value of $\vo$ is needed, resulting in a larger tuning of
$A_s$. We shall therefore focus on the sequestered case
where the string scale turns out to be
$M_s =\sqrt{\pi} g_s^{1/4} M_P/\sqrt{\vo}\simeq 2\cdot 10^{14}$ GeV.
The unification scale is instead set by the winding scale
$M_W \simeq M_s\vo^{1/6}\simeq 5\cdot 10^{15}$ GeV.
The gravitino mass is intermediate,
$m_{3/2} = e^{K_{\rm cs}/2} \sqrt{g_s/2} |W_0| M_P/\vo \simeq 6\cdot 10^{10}$ GeV,
whereas the moduli masses scale as:
\be
m_{\tau_s}\simeq m_{3/2}\ln(M_P/m_{3/2})\simeq 1\cdot 10^{12}\,\text{GeV}\qquad\text{and}\qquad
m_\vo \simeq \frac{m_{3/2}}{\sqrt{\vo}}\simeq 5\cdot 10^6\,\text{GeV}\,,
\ee
showing that there is no cosmological moduli problem. Both
$\tau_{\rm vs}$ and $b$ get a mass of order the string scale,
justifying the fact that they can be integrated out. On the other hand,
the complex structure moduli and the dilaton
develop a mass of order $m_{3/2}$. They can however still be integrated
out at leading order since they are decoupled from
the $T$-moduli (at least at classical level the K\"ahler potential factorises). As far as the axions are concerned,
the $T_s$-axion gets a mass of order $m_{\tau_s}$ while
the volume axion is massless at this level of approximation. It will
only get a tiny mass by subleading non-perturbative effects
which scale as $e^{-\vo^{2/3}}\ll 1$. This bulk axion behaves as dark radiation \cite{1208.3562, 1208.3563} and gets produced by
the decay of the light volume modulus which drives reheating after the end
of inflation.

\section{Summary and conclusions}
\label{sec:Conclusions}

The general ideas regarding realistic local string models and moduli
stabilisation are based on several reasonable assumptions but it is
important to substantiate them with concrete
realisations in order to put those proposals on firmer grounds.
We have been able to combine several
ideas that have been implemented independently before -- chiral matter at branes at singularities, fluxes to stabilise
complex structure moduli and quantum effects to stabilise K\"ahler
moduli -- in one single framework. This is a technically challenging achievement and
represents a clear step forward towards the explicit construction of
fully realistic models. It is very encouraging that 
properties such as the unification and the supersymmetry breaking
scale are obtained at realistic values while at the same time allowing
for de Sitter vacua.

The procedure followed here has been systematic:
\begin{itemize}
\item
First of all, we find appropriate CY manifolds that satisfy the phenomenological requirements: having at least four K\"ahler moduli, two rigid 
cycles mapped into each other by orientifold involution to guarantee a visible sector with $U(n)$ gauge symmetries, an extra rigid cycle in the 
geometric regime for non-perturbative moduli stabilisation and an extra cycle which controls the
overall volume.
Furthermore, these varieties admit suitable discrete symmetries to effectively reduce the number of complex structure moduli.

\item We identify the orientifold action and its associated O-planes. We determine the structure and location of D-branes wrapping different cycles in order to satisfy all consistency conditions. In addition, we make sure that at least one cycle allows a non-zero contribution to the non-perturbative superpotential.

\item In order to stabilise the dilaton and the complex structure moduli, we turn on arbitrary fluxes on three-cycles invariant under the discrete symmetry. This reduces the complex structure stabilisation to a manageable problem.

\item All geometric moduli are stabilised in a self-consistent manner. Assuming that the overall volume is large enough to justify a $1/\mathcal{V}$ expansion, we start with the dominant terms of order $1/\mathcal{V}^2$. These are the F-terms for the dilaton and the complex structure moduli,
    as well as the D-terms. Since all of them are positive definite, they minimise the overall scalar potential at zero value to leading order in $1/\mathcal{V}$. The F-terms fix the dilaton and complex structure moduli. The D-terms induced by anomalous $U(1)$s, both, at the visible and hidden sector, fix the value of matter fields in the hidden sector and blow-up modes in the visible sector. The corresponding axion fields get eaten by the Abelian gauge bosons as in the standard St\"uckelberg mechanism. At next order, i.e. $1/\mathcal{V}^3$, the geometric K\"ahler moduli are stabilised as in the standard LVS scenario, giving rise to an exponentially large volume and supersymmetry breaking.

    Besides the three `standard terms' that determine the large volume minimum, there are non-vanishing F-terms of bulk matter fields residing on the brane wrapping the large cycle. These additional contributions are induced by D-term stabilisation and supersymmetry breaking. They add a positive term which uplifts the value of the minimum. 
    
Depending on the VEV of the flux superpotential $|W_0|$, the minimum can be AdS, dS or Minkowski. We select the value (in the dense uniform range) that gives approximately Minkowski and then let the dense distribution of values around it do the tuning {\it a la} Bousso-Polchinski (although we do not address this part of the framework here). 

Depending on the structure of the induced soft terms in the visible sector, the corresponding matter fields get a zero or non-zero VEV. In turn, the visible sector blow-up mode develops a VEV which can be either vanishing or suppressed by inverse powers of $\vo$, justifying the validity of working in the singularity regime.

\item The scales determined by this stabilisation procedure acquire realistic values, in the sense that the string scale is close to the standard 
GUT scale, while the soft terms are either of the order of the TeV-scale (fitting with a sequestered scenario) or of the order of the gravitino mass 
which is heavy (intermediate scale). In both scenarios the lightest volume modulus is heavier ($10^6-10^7$ GeV) than the TeV scale, e.g.~the 
gaugino mass scale in the sequestered case. Therefore, there is no cosmological moduli problem and the decay of this mode determines 
reheating dominating late time cosmology.

\item In the case with a non-vanishing VEV for a visible sector matter field, its phase becomes a potential candidate for the QCD axion with a 
decay constant in the allowed range and similar to the (large) value of the gravitino mass $\sim 10^{10}$ GeV. The detailed structure of VEVs for 
matter fields is more model-dependent, depending in part on the distribution of D3-branes at the dP$_6$ singularity. Since there are many 
options we will leave this for a future treatment, as well as a discussion of the phenomenology of this class of models.
\end{itemize}


It is interesting to note that the values for $|W_0|$ found in our analysis are concentrated within the range 
$1 \lesssim |W_0|\lesssim 100$.\footnote{Notice that for technical reasons, we have turned on only a small number of flux quanta. Very small 
values of $|W_0|$, as required for KKLT, are expected to emerge only when hundreds of flux quanta are switched on.} This is consistent with the 
following two important requirements necessary to trust the 4D low-energy theory. Firstly, we must have 
$\vo \gg |W_0|^3$ in order to control the superspace derivative expansion by having a gravitino mass smaller than the Kaluza-Klein scale 
$M_{\rm KK}$ \cite{Cicoli:2013swa}. Secondly, we need $\vo \gg |W_0|^6$ to have a vacuum energy density smaller than $M_{\rm KK}^4$ 
\cite{Conlon:2005ki,Cicoli:2013swa,deAlwis:2012vp}. 
However, the last constraint applies only to AdS vacua, and so it does not affect our Minkowski or slightly de Sitter solutions, whereas
the first constraint is more generic. For $|W_0|\simeq 100$ it requires a volume of order $\vo \gg 10^6$. This bound can be easily satisfied in 
the LVS scenario due to the presence of an exponentially large volume. In the explicit minimisation carried out in Section \ref{sec:KaehlerStab}, 
we actually found a minimum with $\vo \simeq 10^8$. 

There are many open questions to be addressed. In order to incorporate 
cosmological issues into this framework, it would be interesting to considering concrete CY
manifolds which allow for a realisation of inflation. Two of the models in Appendix~\ref{app:MoreEx} have $h^{1,1}=5$ and are K3-fibred. 
These are the right features to realise different promising models where the inflaton is a K\"ahler 
modulus~\cite{Conlon:2005jm,Bond:2006nc,Cicoli:2008gp,Cicoli:2011ct}. The moduli space of the open string sector needs to be studied in 
order to combine moduli stabilisation with a realistic higgsing towards the
Standard Model gauge group. 

Moreover, in this paper we turned on flux quanta only along three-cycles which are invariant under an appropriate discrete symmetry of the 
complex structure moduli space. Even in this simpler case, we could find only a few solutions to the minimisation equations which are 
completely under control, in the sense that the instanton corrections determine only a negligible shift of the VEV of the $u$-moduli. More 
generically, in order to perform a proper statistical analysis of flux vacua, to understand the distribution of phenomenologically interesting mass 
scales in the landscape, one should turn on flux quanta on all three-cycles. In addition one needs to be able to find solutions to the complete 
minimisation equations including also instanton effects. This task is a real technical challenge that we leave for future work. However, we still 
think that our paper represents an important step forward in the understanding of crucial questions, like for example the explicit realisation of 
the Bousso-Polchinski scenario for the cosmological constant~\cite{Bousso:2000xa}.

\section*{Acknowledgements}

We would like to thank Martin Bies, Mirjam Cveti\v{c}, Andres Collinucci,
Jim Halverson, Albrecht Klemm, Francesco Muia and Markus Rummel
for useful conversations.
The research of DK is supported in part by
the DOE grant DE-SC0007901 and by Dean's Funds for Faculty Working
Group.
The work of SK and CM was supported by the DFG under TR33 ``The Dark
Universe''. SK was also supported by the European Union 7th network
program Unification in the LHC era (PITN-GA-2009-237920). SK and  MC
would like to thank ICTP for hospitality.

\appendix

\section{Periods and Picard-Fuchs equations}
\label{app:Review}

In this appendix we review some standard facts about mirror symmetry, special geometry and the PF-system that are omitted in
the main text for brevity of the discussion, but might prove helpful for the interested reader. We also present
explicit formulas for the period vector $\Pi$ and the flux superpotential $W_{\rm flux}$ in terms of the prepotential
$F$. For more details we refer to standard references, e.g.~\cite{Marino:2004uf,Klemm:2005tw,Vonk:2005yv}. We closely follow 
\cite{Klevers:2011xs}.

\subsection{Periods as solutions to the Picard-Fuchs equations}
\label{sec:PFOtoricHS}

In this section we collect some derivations and explanations that have been skipped in the main text. We
will be as brief as possible and refer to the context of the main text frequently, in order to avoid repetitions.

Next we discuss the structure of the complex structure moduli space $\mathcal{M}_{\rm cs}$ of $M_3$.  This
structure is captured by the dependence of the holomorphic three-form $\Omega$ and its
periods $\Pi$ on the complex structure moduli $\underline{z}$. The periods are solutions
to the PF differential system. In the following we are mostly interested in the periods
at the point of large complex structure $\underline{z}\sim 0$. Invoking mirror symmetry, the
periods in the symplectic basis of three-cycles on $M_3$ are determined by classical intersections on
$\tilde{M}_3$. In turn worldsheet-instanton corrections to the K\"ahler moduli space of $\tilde{M}_3$ are predicted
by the expansion of the periods of $M_3$ at large complex structure/large volume.

The complex structure dependence of the holomorphic three-form $\Omega$ is conveniently parametrized by its periods,
introduced in the period expansion \eqref{eq:periodExp} w.r.t.~an integral symplectic basis $(\alpha_K,\beta^K)$ of
$H^3(M_3,\Z)$. The basis elements  obey the elementary relations
\beq
	\int \alpha_K\wedge\beta^L=-\int \beta^L\wedge\alpha_K=\delta_K^L\,,\quad \int \alpha_K\wedge\alpha_L=-\int \beta^L\wedge\beta^K=0\,.
\eeq
In terms of the Poincar\'e dual basis $(A_K,B^L)$ of $H_3(M_3,
\mathbb{Z})$ defined via
\beq
	\int_{A_K}\alpha_L=-\int_{B^L}\beta^K=\delta^{K}_L
\eeq
the periods of $\Omega$ can be directly defined
by the following integrals
\beq
	X^K=\int_{A_K}\Omega\,,\qquad \mathcal{F}_K=\int_{B^K}\Omega\,.
\eeq
The $X^K$ form projective coordinates on the complex structure moduli space since
$\Omega$ is only defined up to multiplication with $e^f$ for a
holomorphic function $f$ on the complex structure moduli space\footnote{More precisely $\Omega$ is a section of the
vacuum line bundle $\mathcal{L}$ over $\mathcal{M}_{\rm cs}$, see \cite{Hori:2003ic}.}. Mathematically
the existence of these coordinates reflects the isomorphism $H^1(M_3,TM_3)\cong H^{(2,1)}(M_3)$
and the local Torelli theorem. Thus the dimension of complex structure moduli space is $h^{(2,1)}$ and the periods
$\underline{\mathcal{F}}$ have to be functions of the $\underline{X}$. Employing Griffiths transversality \cite{Griffiths:1969fr,Griffiths:1969frII} in the form
\begin{equation}
	0=\int_{M_3}\Omega\wedge \frac{\partial}{\partial X^I}\Omega\,.
\label{eq:GriffithsTransversalityRelation}
\end{equation}
it can be shown that  $\mathcal{F}_{K}=X^L\frac{\partial}{\partial X^K}\mathcal{F}_L=\frac{\partial}{\partial X^K}(X^L\mathcal{F}_L)-\mathcal{F}_I$, which immediately implies \eqref{eq:prepotProjective}. The gauge freedom can be
fixed by setting  $e^f=1/X^0$ yielding \eqref{eq:prepotAffine} and \eqref{eq:flatcoord}.

Next, we note that the natural hermitian metric
\begin{equation}
	h=i\int\Omega\wedge\bar\Omega\,.
\label{eq:HermitMetricL}
\end{equation}
on the vacuum line bundle $\mathcal{L}$ spanned by $\Omega$
\cite{Hori:2003ic} also enters the K\"ahler potential \eqref{eq:KaehlerpotCS} on the complex structure moduli space.
Similarly $h$ is used to construct the hermitian/Chern connection on
the holomorphic line bundle $\mathcal{L}$,
\begin{equation}
		A_i=\partial_i\log(h)=-\partial_i K_{\rm cs}=\frac{i}{h}(\bar{X}^K\partial_i\mathcal{F}_K-\bar{\mathcal{F}}_{\bar i})\,.
\label{eq:Connection+CurvatureL}
\end{equation}
From this one can determine the covariant derivative on $\cL$ as $D_i=\partial_i+\partial_iK_{\text{cs}}$, which has to be
used e.g.~when evaluating the F-terms \eqref{eq:F-term}.
Theses definitions imply that the  curvature form of the bundle $\mathcal{L}$ is also the K\"ahler form on its base
manifold, i.e.~the complex structure moduli space. This is one central property of $\mathcal{N}=2$ special geometry.

As mentioned in the main text, the main objective in order to control the complex structure sector of a CY three-fold $M_3$
is to compute the prepotential \eqref{eq:prepotAffine}. This is accomplished by solving the associated PF-system, which
is constructed as follows.
For toric hypersurface the explicit algebraic representation of a family of CY hypersurfaces  \cite{Batyrev:1994hm}  in a toric variety
$\mathds{P}_{\Delta}$ associated  to $\Delta_4^Z$  reads
\begin{equation} \label{eq:Z3typeIIB}
 	 P(\underline{x},\underline{a})=\sum_{p_j\in \Delta^{\tilde Z}_4}a_j\prod_{i}x_i^{\langle v_i,p\rangle+1}\,.
\end{equation}
Here, we introduced the projective coordinates $x_i$ to each vertex
$v_i$ of $\Delta_4^Z$ and $p_j$ are the integral points in the dual polytope $\Delta_{4}^{\tilde{M}}$. The constants
$a_j$ are labeled by these points and constitute the complex structure moduli of $M_3$.
It can be shown, using a residue integral representation for $\Omega$ \cite{Griffiths:1969fr,Griffiths:1969frII,Griffiths:1968,Griffiths:1968II,griffiths1970periods,griffiths1970summary}, that directly involve
the constraint \eqref{eq:Z3typeIIB}, that $\Omega$ obeys the Gelfand-Kapranov-Zelevinski (GKZ)
differential system \cite{gelfand1989hypergeometric,gelfand1990generalized}.
This system is constituted of the operators ${\cal L}_l$ and ${\cal Z}_k$. They are completely
determined by the toric data encoded in the points $\tilde{v}_j$ and $\ell^{(i)}$-vectors $\ell^{(a)}$
of $\Delta_4^{\tilde{M}}$ as
\begin{equation} \label{eq:pfo}
 	\mathcal{L}_{i}=\prod_{\ell^{(i)}_j>0}\left(\frac{\partial}{\partial a_j}\right)^{\ell^{(i)}_j}
 	-\prod_{\ell^{(i)}_j<0}\left(\frac{\partial}{\partial a_j}\right)^{-\ell^{(i)}_j}\,,\qquad i=1,\ldots, k\,,
\end{equation}
that are the actual PF operators annihilating the periods and operators
\begin{equation} \label{eq:Zs}
 	\mathcal{Z}_{j}=\sum_n(\bar{v}_n)^j\vartheta_n-\beta_j\,,\qquad j=0,\ldots,4\,.
\end{equation}
Here $\beta=(-1,0,0,0,0)$ is the so-called exponent of the GKZ-system and
$\vartheta_n=a_n\frac{\partial}{\partial a_n}$ denote logarithmic derivatives. We have
embedded the points $\tilde{v}_n$ into a hypersurface at distance $1$ away from the origin by
defining $\bar{v}_n=(1,\tilde{v}_n)$ so that all zeroth components are $(\bar{v}_n)^0=1$.

The differential equations \eqref{eq:Zs} are easily solved by introducing coordinates
\begin{equation} \label{eq:algCoords}
 	z^i=(-)^{\ell^{(i)}_0}\prod_{j=0}^{k+4} a_j^{\ell^{(i)}_j}\,,\quad i=1,\ldots k\,,
\end{equation}
on the complex structure moduli space of $M_3$.
In these so-called algebraic coordinates the point of maximally unipotent monodromy, also denoted as the point of large volume/large
complex structure, is centred at $z^i=0$.
Indeed by solving  the GKZ-system in these coordinates
explicitly using the Frobenius method \cite{Hosono:1993qy,Hosono:1994ax} the obtained solutions have the well-known
log-grading.
There is a  a unique holomorphic power series solution $X^0$ and solutions starting with $(\ln(z^i))^k$ for $k=0,\ldots, 3$.
This structure is expected by mirror symmetry due to the shift symmetry of the NS-NS B-field, $b^i\mapsto b^i+1$, in the Type IIA
theory, which implies on the B-model side a maximal logarithmic degeneration of periods near the point $z^i=0$
\cite{Candelas:1990rm,Hosono:1993qy}.

In order to relate these solutions with the geometrical periods $\Pi$ in \eqref{eq:PiXF} of $\Omega$ in
the integral symplectic basis $(\alpha_K,\beta^K)$ one employs mirror symmetry.
The mirror map \eqref{eq:t^iIIA} is constructed by identifying the $X^i$ in \eqref{eq:flatcoord} with the
single-logarithmic solutions in the $z^i$. Then, we have to find the prepotential \eqref{eq:pre_largeV} among the
solutions of the PF system, which also fixes all other periods by  \eqref{eq:PiXF}.
It can be argued invoking the matching of D-brane central charges under
mirror symmetry \cite{Mayr:2000as} that, in a suitable integral basis, the
subleading terms of $F$ are given by the classical intersections\footnote{We note that the
last period in \eqref{eq:PiXF} only matches the vanishing period at
the conifold of the mirror quintic \cite{Candelas:1990rm} if $\mathcal{K}_{ij}$
is shifted by $8$, which amounts to replace the integrand of $\cK_{ij}$ by
$(J^2+c_2(M_3))$ with $c_2(M_3)$ denoting the second Chern class of
$\tilde{M}_3$ \cite{Huang:2006hq}.} \eqref{eq:classic_terms_threefold2} of the mirror
three-fold $\tilde M_3$, however, with
\beq \label{eq:KijMayr}
  \cK_{ij}=\frac{1}{2}\int_{\tilde M_3} \iota_*(c_1(K_j)) \wedge J_i \,\,\,\,\text{mod}\, \mathbb{Z}\,.
\eeq
Here $c_1(K_j)$ denotes the first Chern class of the divisor $K_j$ and
$\iota_*=P_{\tilde M_3}\circ\iota_*\circ P^{-1}_{K_j}$ is the Gysin
homomorphism where $P_{\tilde M_3}$
($P_{\tilde K_j}$) is the Poincar\'e-duality map on $\tilde M_3$ ($K_j$) and
$\iota_*$ is the push-forward
on homology of the embedding $\iota:\, K_j\rightarrow \tilde{M}_3$. Thus,
$\iota_*(c_1(K_j))$ is a four-form. It is calculated by first noting the
projection formula \cite{fulton1984intersection} for a map $f:\,X\rightarrow X'$ between two
algebraic varieties $X$, $X'$,
\beq
	f_*([Y]\cdot f^*([Y']))=f_*([Y])\cdot [Y']
\eeq
for two subvarieties $Y$, $Y'$ of $X$ respectively $X'$ with corresponding
homology classes $[Y]$, $[Y']$. Applying this
to the embedding $\iota$ we obtain
\beq
	\iota_*(c_1(K_j))\cdot J_i=\iota_*(c_1(K_j)\cdot
	\iota^*(J_i))=P_{\tilde{M}_3}([pt])\int_{K_j}c_1(K_j)\iota^*(J_i)
\eeq
where Poincar\'e duality in the first and second equation is understood. Here
we have used that on $K_j$ any four-form is proportional to the top form on
$K_j$ with proportionality factor given by its integral over $K_j$. Using the adjunction formula for $K_j$ in $\tilde{M}_3$
to show $c_1(K_j)=-\iota^*J_j$, we finally obtain
\beq
	\mathcal{K}_{ij}=-\frac{1}{2}\int_{\tilde{M}_3}
	J_i\wedge J_j^2\,\,\,\,\text{mod}\, \mathbb{Z}\,,
\eeq
which is precisely the expression in \eqref{eq:classic_terms_threefold2}.

For reference in the main text, we conclude this appendix by presenting the period vector \eqref{eq:PiXF}
in the integral basis that follows from the prepotential \eqref{eq:pre_largeV}:
\begin{equation}
	\Pi=X^0\begin{pmatrix}
				1 \\
				u^i\\
								\tfrac{1}{3!} \cK_{ijk}\, u^i u^j u^k  + \cK_{i} u^i + \cK_0 + \sum_{\beta} n_\beta^0\,(2\text{Li}_3(q^\beta)-d_iu^i\text{Li}_2(q^\beta))\\
				-\tfrac{1}{2} \cK_{ijk}\, u^j u^k - \cK_{ij}\, u^j + \cK_{i} + \sum_{\beta} n_\beta^0d_i\,\text{Li}_2(q^\beta)
				\end{pmatrix}\,.
\label{eq:periodVectorCY3LV}
\end{equation}
We also display  flux superpotentials obtained for this period vector
\begin{eqnarray} \label{eq:flux_ori}
   W_{\rm flux} \!\!&\!\!=&\!\!\! X^0\big[\hat N_0 + \hat{M}^0\mathcal{K}_0+\hat{M}^i\mathcal{K}_i+(\hat N_i-\hat{M}^j\mathcal{K}_{ij}+\hat{M}^0\mathcal{K}_i) u^i
   -  \tfrac12\hat M^i \cK_{ijk}\, u^j u^k \nn\\
   \!\!&\!\!+&\!\!\!\tfrac1{3!}\hat{M}^0\mathcal{K}_{ijk}\,u^i u^j u^k+(\hat{M}^i-\hat{M}^0 u^i)\sum_{\beta} d_i n_\beta^0\,\text{Li}_2(q^\beta)
   +2\hat{M}^0\sum_{\beta} n_\beta^0\,\text{Li}_3(q^\beta)\big]\,
\end{eqnarray}
where the flux $G_3=F_3-\tau H_3$ has been expanded into complex flux quantum numbers
$(\hat M_K,\hat N^K) = (M_K -\tau \tilde M_K,N^K - \tau \tilde N^K)$ formed from the flux quantum numbers $(M_K,N^K)$ of $F_3$ and
$(\tilde M_K,\tilde N^K)$ of $H_3$.

\section{More examples}
\label{app:MoreEx}

In this section we calculate the period vector at large complex structure for
selected examples of CY three-folds $X_3$ which satisfy the phenomenological criteria established in~\cite{Cicoli:2012vw}. These CY three-folds
are realized as hypersurfaces in toric varieties. The number of complex structure moduli
$\underline{z}$ in the examples ranges between four and five.

\subsection{Example 2}
\label{sec:example2}

Here we consider another four moduli example of mirror pairs of
CY three-folds $(X_3,\tilde{X}_3)$. The polytope $\Delta_4^X$ and
its dual $\Delta_4^{\tilde{X}}$ read
\beq
	\Delta_4^{X} =
	\left(
	\begin{array}{rrrr||c}
		2& 1&-1&-1& D_1\\
		2& 1& 0&-2& D_2\\
		2& 2&-1&-2& D_3\\
		2& 2& 0&-3& D_4\\
		0& 0& 0& 1& D_5\\
		0& 0& 1& 0& D_6\\
        0& 1& 0& 0& D_7\\
       -1&-1& 0& 1& D_8
	\end{array}
	\right),\qquad
	\Delta_4^{\tilde{X}} =
	\left(
	\begin{array}{rrrr||c}
	-1&-1&-1&-1& \tilde{D}_1\\
	 1&-1&-1&-1& \tilde{D}_2\\
     1&-1& 3&-1&\tilde{D}_3\\
	-3& 3&-1&-1& \tilde{D}_4\\
	 5&-1&-1& 3& \tilde{D}_5\\
	 5&-1& 3& 3& \tilde{D}_6\\
	 1& 3&-1& 3& \tilde{D}_7\\
	 1& 3& 3& 3& \tilde{D}_8
\end{array}
	\right)\,.
	\label{eq:delta4ex2}
\eeq
Here we have introduced the toric divisors $D_i$ respectively
$\tilde{D}_i$ corresponding to the vertices in $\Delta_{4}^{X}$
respectively $\Delta_{4}^{\tilde{X}}$ in the last column.
We note that the polytopes $\Delta_4^{X}$ and $\Delta_4^{\tilde{X}}$ are
congruent. Thus the corresponding toric varieties $\mathds{P}_{\Delta}$ and
$\mathds{P}_{\tilde{\Delta}}$ differ only by the action of an orbifold group and
the mirror construction agrees with the GP orbifold
construction of the mirror CY three-fold.

We note again that it is the CY three-fold $X_3$ of
which we want to calculate the periods on its complex structure moduli
space. Its Hodge numbers and Euler number read
\beq
	h^{(1,1)}=4\,,\quad h^{(2,1)}=98\,,\quad \chi(X_3)=-188\,.
\eeq
After demanding invariance under the GP orbifold group,
only four of the $98$ complex structure moduli remain. The periods
left invariant under this orbifold precisely agree with the periods
of the mirror $\tilde{X}_3$, as explained in Section \ref{sec:ReducingCSM}, that we calculate next.

All the following geometrical calculations are performed on the K\"ahler
moduli space of $X_3$.
First we note that there are 16 star-triangulations of $\Delta_4^{X}$.
By calculating the quartic intersections on the toric ambient variety we see
that there are only eight CY phases on $X_3$, both with simplicial
and non-simplicial Mori cone. We focus in the following on a CY phase with simplicial
Mori cone that arises from a single phase of the ambient toric variety.
The Mori cone in this phase is generated by
\bea \label{eq:ellsZ3ex2}
	&\ell^{(1)}=(-1,1,1,-1,0,0,0,0)\,,&\qquad \ell^{(2)}=
	(0,0,0,1,1,0,0,2)\,,\\
&\ell^{(3)}=(1,-1,0,0,-1,1,0,0,)\,,&\qquad
\ell^{(4)}=(1,0,-1,0,-1,0,1,0)\nn
\eea
Next, we readily calculate by duality between curves and divisor,
\beq
	K_i\cdot \ell^{(j)}=\delta^{j}_i\,,
\eeq
the generators $K_i$ of the K\"ahler cone. In the chosen
triangulation the K\"ahler cone is spanned by the generators $K_i$
\beq \label{eq:Kaehlerconeex2}
	K_1=D_1+D_2+D_3\,,\quad K_2=D_1+D_2+D_3+D_4 \,,\quad K_3=D_1+D_3\,,\quad
	K_4=D_1+D_2
\eeq
in terms of the toric divisors $D_i$ introduced in \eqref{eq:delta4ex2}.
In terms of this basis we calculate the classical triple intersections w.r.t.~the $(1,1)$-forms $J_i$ dual to the $K_j$ as
\bea \label{eq:C0ex2}
	\mathcal{C}_0&=&J_1^2 J_2+3 J_1 J_2^2+\frac{5 J_2^3}{3}+2 J_1 J_2 J_3+2 J_2^2 J_3+2 J_1 J_2 J_4+2 J_2^2 J_4+2 J_2 J_3 J_4\,.
\eea
Analogously we calculate the
other classical intersections \eqref{eq:classic_terms_threefold2} to be
\bea \label{eq:subleadingtermsex2}
	\cK_{12}= -3\,,\,\, \cK_{21}= -1\,,\,\,\cK_{32}=\cK_{42}=-2\,,\,\,\,\,
	 \cK_{22}=-5\,, \\
  \cK_j J_j = \frac{2}{3}J_1+\frac{8}{3}J_2+J_3+J_4\ , \qquad
  \cK_0= -\frac{\zeta(3)}{(2 \pi i)^3}188\,
\eea
with all other $\cK_{ij}=0$. We note that these relations hold up to integers
as before and the $\cK_{ij}$ are symmetric modulo integers.

Finally, the intersections \eqref{eq:C0ex2} and
\eqref{eq:subleadingtermsex2} allow us to write down the prepotential
\eqref{eq:pre_largeV} as
\bea
	F&=&-t_1^2 t_2-3 t_1 t_2^2-\frac{5 t_2^3}{3}-2 t_1 t_2 t_3-2 t_2^2 t_3-2 t_1 t_2 t_4-2 t_2^2 t_4-2 t_2 t_3 t_4\nn\\&&
 +2 t_1t_2+t_2t_3+t_2t_4+\frac{5}{2}t_2^2
 +\frac{2}{3}t_1+\frac{8}{3}t_2+t_3+t_4-i\zeta(3)\frac{47}{4\pi^3}\nn\\
 &&
 +  \sum_{\beta} n_\beta^0\,\text{Li}_3(q^\beta)\,,
\eea
where $\beta=(d_1,d_2,d_3,d_4)$ in the basis $\beta^i$ of effective curves
of $H_2(X_3,\mathbf{Z})$ corresponding to the generators
\eqref{eq:ellsZ3ex2} of the Mori cone and $q^\beta=e^{2\pi id_it^i}$ as
before. We also introduced the flat coordinates $t_i$ on the K\"ahler
moduli space of $X_3$ corresponding to the $\beta^i$. The
instanton corrections for the K\"ahler moduli space are determined by
solving the PF equations for $\tilde{X}_3$ along the lines of section
\ref{sec:PFOtoricHS} for the charge vectors \eqref{eq:ellsZ3}.

\subsubsection{The orientifold involution}

The weight matrix for $\Delta_4^{X}$, which was given in \eqref{eq:delta4ex2}, reads as follows:
\begin{equation}
\begin{array}{|c|c|c|c|c|c|c|c||c|}
\hline z_1 & z_2 & z_3 & z_4 & z_5 & z_6 & z_7 & z_8 & D_{eq_X} \tabularnewline \hline \hline
1 &  0 &  0 &  1 &  0 &  1 &  1 &  4 & 8\tabularnewline\hline
0 &  1 &  0 &  0 &  0 &  0 &  1 &  2 & 4\tabularnewline\hline
0 &  0 &  1 &  0 &  0 &  1 &  0 &  2 & 4\tabularnewline\hline
0 &  0 &  0 &  1 &  1 &  0 &  0 &  2 & 4\tabularnewline\hline
\end{array}
\label{eq:ex2:weightm}\, .
\end{equation}
Again we note that this phase differs from the one used in \eqref{eq:ellsZ3ex2}.
The Stanley-Reisner ideal in this phase reads:
\begin{equation}
\textmd{SR-ideal}:\qquad\{z_1 \, z_6 ,\, z_1 \, z_7 ,\, z_2 \, z_3 ,\, z_2 \, z_7 ,\, z_3 \, z_6 ,\, z_4 \, z_5 \, z_8\}
\end{equation}
The hypersurface equation at the symmetric complex structure points is given by:
\begin{equation}
\begin{aligned}
&\textmd{equ}_\textmd{rd}=z_1^4 \, z_2^4 \, z_3^4 \, z_4^4 + z_2^4 \, z_4^4 \, z_6^4 + z_1^4 \, z_2^4 \, z_5^4 \, z_6^4 +
z_3^4 \, z_4^4 \, z_7^4 + z_8^2 + \rho_1\,z_1^8 \, z_2^4 \, z_3^4 \, z_5^4  +\\& \quad+
\rho_2\,z_1^4 \, z_3^4 \, z_5^4 \, z_7^4  + \rho_3\,z_5^4 \, z_6^4 \, z_7^4  +
\rho_4\, z_1 \, z_2 \, z_3 \, z_4 \, z_5 \, z_6 \, z_7 \, z_8 +
\rho_4^2\,z_1^2 \, z_2^2 \, z_3^2 \, z_4^2 \, z_5^2 \, z_6^2 \, z_7^2
\end{aligned}
\end{equation}

The holomorphic involution is given by:
\begin{equation}
 \sigma:\quad (z_2\,,\,z_6)\leftrightarrow(z_3\,,\,z_7)\,.
\end{equation}
There is only one $O7$-plane and it is given by:
\begin{equation}
 O7:\quad z_2\,z_6-z_3\,z_7=0\,.
\end{equation}
with $\chi(O7)=36$. Furthermore, we have $O3$-planes at
\begin{equation}
 \{z_1= 0,\, z_5= 0,\, z_2\,{z_6}+z_3\,z_7=0\}\quad\textmd{and}\quad  \{z_1=0,\, z_4=0,\,  z_2\,{z_6}+z_3\,z_7=0\}
\end{equation}

\subsection{Example 3}
\label{sec:example3}

In this section we consider a five moduli example of mirror pairs of
CY three-folds $(X_3,\tilde{X}_3)$. The polytope $\Delta_4^X$ and its dual $\Delta_4^{\tilde{X}}$ read
\beq
	\Delta_4^{X} =
	\left(
	\begin{array}{rrrr||c}
		3& 1&-1&1& D_1\\
		6&-1& 0& 2& D_2\\
		0& 1& 0& 0& D_3\\
		6&-2& 1& 2& D_4\\
		0& 0& 1& 0& D_5\\
	    0& 0& 0& 1& D_6\\
       -1& 0& 0&-1& D_7\\
        3& 0& 0& 1& D_8\\
        3&-1& 1& 1&D_9\\
	\end{array}
	\right),\qquad
	\Delta_4^{\tilde{X}} =
	\left(
	\begin{array}{rrrr||c}
	-1&-1&-1& 2& \tilde{D}_1\\
	 0&-1&-1&-1&\tilde{D}_2\\
     2&-1&-1&-1&\tilde{D}_3\\
	 2&-1& 5&-1& \tilde{D}_4\\
	 2& 5&-1&-1& \tilde{D}_5\\
	 2&11&11&-1&\tilde{D}_6\\
	 2&11&17&-1& \tilde{D}_7\\
\end{array}
	\right)\,.
	\label{eq:delta4ex3}
\eeq
Here we have introduced the toric divisors $D_i$ respectively
$\tilde{D}_i$ corresponding to the points in $\Delta_{4}^{X}$
respectively $\Delta_{4}^{\tilde{X}}$ in the last column. We have included
two interior points on codimension two and three faces in $\Delta_{4}^{X}$
as the last two points in \eqref{eq:delta4ex3}. These points are necessary
to resolves singularities of the toric ambient space which also descend to the generic CY
hypersurface $X_3$.
We note that the polytopes $\Delta_4^{X}$ and $\Delta_4^{\tilde{X}}$ are
congruent.

The Hodge numbers and Euler number of the CY three-fold $X_3$ of
which we want to calculate the periods read
\beq
	h^{(1,1)}=5\,, \quad h^{(2,1)}=185\, 
\,,\quad \chi(X_3)=-360\,.
\eeq
After demanding invariance under the GP orbifold group,
only five of the $185$ complex structure moduli remain. The periods
left invariant under this orbifold precisely agree with the periods
of the mirror $\tilde{X}_3$, that we calculate in the next.

All the following calculations are performed on the K\"ahler
moduli space of $X_3$. We note that the ambient toric variety $P_{\Delta}$
has 13 inequivalent star-triangulations. As before we calculate the quartic intersections
of its toric divisors, noting that all these 13 phases descend to distinct CY phase
of $X_3$.
We focus here on a particular star-triangulation of $\Delta_{4}^{X}$ with simplicial Mori cone generated by
\bea \label{eq:ellsZ3ex3}
	&\ell^{(1)}=(0, 0, -1, -1, 0, 0, 0, 1, 1)\,,\, \ell^{(2)}=
	(0, 0, 0, 1, 1, 0, 0, 0, -2)\,,\,\ell^{(3)}=(0, 0, 1, 0, -1, 2, 3, 0, 1),&\nn\\
	&\ell^{(4)}=(0, 1, 1, 0, 0, 0, 0, -2, 0)\,,\quad \ell^{(5)}=(1, -1, 0, 1, 0, 0, 0, -1, 0)\,.&
\eea
Next, we readily calculate by duality between Mori and K\"ahler cone,
the generators $K_i$ of the K\"ahler cone. In the chosen
triangulation the K\"ahler cone is spanned by the generators $K_i$
\bea \label{eq:Kaehlerconeex3}
	&K_1=D_1-D_4+D_5+\tfrac{1}{2}D_6\,,\quad D_2=D_5+
	\tfrac{1}{2}D_6\,,\quad K_3=\tfrac{1}{2}D_6&\nn\\
	&K_4=D_1+D_3-D_4+D_5\,,\quad K_5=D_1&
\eea
in terms of the toric divisors $D_i$ introduced in \eqref{eq:delta4ex2}.
In terms of this basis we calculate the classical triple intersections as
\bea \label{eq:C0ex3}
	\mathcal{C}_0\!\!&\!\!=\!\!&\!\!\frac{4 J_1^3}{3}+4 J_1^2 J_2+4 J_1 J_2^2+\frac{7 J_2^3}{6}+4 J_1^2 J_3+8 J_1 J_2 J_3+\frac{7}{2} J_2^2 J_3+4 J_1 J_3^2+\frac{7}{2} J_2 J_3^2+J_3^3\\
	&+&\frac{3}{2} J_1^2 J_4+3 J_1 J_2 J_4+\frac{3}{2} J_2^2 J_4+3 J_1 J_3 J_4+3 J_2 J_3 J_4+\frac{3}{2} J_3^2 J_4+\frac{1}{2} J_1 J_4^2+\frac{1}{2} J_2 J_4^2+\frac{1}{2} J_3 J_4^2\nn\\
	&+&J_1^2 J_5+2 J_1 J_2 J_5+J_2^2 J_5+2 J_1 J_3 J_5+2 J_2 J_3 J_5+J_3^2 J_5+J_1 J_4 J_5+J_2 J_4 J_5+J_3 J_4 J_5\,.\nn
\eea
Similarly we calculate the
other classical intersections \eqref{eq:classic_terms_threefold2} to be
\bea \label{eq:subleadingtermsex3}
	&\cK_{11}=\cK_{12}=\cK_{13}=\cK_{21}=\cK_{31}=-4\,,\,\, \cK_{14}=\cK_{24}=\cK_{34}= -\frac{1}{2}\,,
	\quad\cK_{22}=\cK_{23}=\cK_{32}=-\frac{7}{2}\,,&\nn\\
	 &\cK_{33}=-3\,,\quad \cK_{41}=\cK_{42}=\cK_{43}=-\frac{3}{2}\,,\quad
	\cK_{51}=\cK_{52}=\cK_{53}=-1\,,&\nn\\
  &\cK_j J_j = \frac{23}{6}J_1+\frac{41}{12}J_2+3J_3+\frac{3}{2}J_4+J_5\ , \qquad
  \cK_0= -\frac{\zeta(3)}{(2 \pi i)^3}360\,&
\eea
with all other $\cK_{ij}=0$. We note that these relations hold up to integers
as before and the $\cK_{ij}$ are symmetric modulo integers.

Finally, the intersections \eqref{eq:C0ex3} and
\eqref{eq:subleadingtermsex3} allow us to obtain the prepotential
\eqref{eq:pre_largeV}. For brevity, we do not write out the cubic term by noting that they
are identical to \eqref{eq:C0ex3} after replacing $J_i\rightarrow -t_i$. We obtain
\bea
	F&=&C_0\vert_{J_i=-t_i}
	+2t_1^2+4t_1t_2+4t_1t_3+\frac{1}{4}t_1t_4+\frac{1}{4}t_2t_4+\frac{1}{4}t_3t_4+\frac{7}{4}t_2^2+\frac{7}{2}t_2t_3
	+\frac{3}{2}t_3^2\nn\\
	&&+t_1t_4+t_2t_4+t_3t_4+\frac{1}{2}t_1t_5+\frac{1}{2}t_2t_5+\frac{1}{2}t_3t_5+\frac{23}{6}t_1
	+\frac{41}{12}t_2+3t_3+\frac{3}{2}t_4+t_5\nn\\
 &&
-i\zeta(3)\frac{45}{2\pi^3} +  \sum_{\beta} n_\beta^0\,\text{Li}_3(q^\beta)\,,
\eea
where $\beta=(d_1,d_2,d_3,d_4)$ in the basis $\beta^i$ of effective curves
of $H_2(X_3,\mathbf{Z})$ corresponding to the generators
\eqref{eq:ellsZ3ex3} of the Mori cone and $q^\beta=e^{2\pi id_it^i}$ as
before. We also introduced the flat coordinates $t_i$ on the K\"ahler
moduli space of $X_3$ corresponding to the $\beta^i$. The
instanton corrections for the K\"ahler moduli space are determined by
solving the PF equations for $\tilde{X}_3$ along the lines of section
\ref{sec:PFOtoricHS} for the charge vectors \eqref{eq:ellsZ3}.

\subsubsection{The orientifold involution}

The weight matrix for $\Delta_4^{\tilde{X}}$, which was given in \eqref{eq:delta4ex3}, reads as follows:
\begin{equation}
\begin{array}{|c|c|c|c|c|c|c|c|c||c|}
\hline z_1 & z_2 & z_3 & z_4 & z_5 & z_6 & z_7 & z_8 & z_9 & D_{eq_X} \tabularnewline \hline \hline
1 &  1 &  0 &  0 &  1 &  6 &  9 &  0 &  0 & 18\tabularnewline\hline
1 &  0 &  1 &  1 &  0 &  6 &  9 &  0 &  0 & 18\tabularnewline\hline
1 &  0 &  0 &  0 &  0 &  4 &  6 &  0 &  1 & 12\tabularnewline\hline
0 &  1 &  1 &  0 &  0 &  4 &  6 &  0 &  0 & 12\tabularnewline\hline
0 &  0 &  0 &  0 &  0 &  2 &  3 &  1 &  0 & 6\tabularnewline\hline\end{array}
\label{eq:ex3:weightm}\, .
\end{equation}
The Stanley-Reisner ideal for the phase of the ambient space, which gives rise to two dP$_8$ divisors and one dP$_1$ divisor on the CY and respects the K3-fibration structure of the polytope, is given by:
\begin{equation}
\begin{aligned}
\textmd{SR-ideal}:\quad\{&
z_1\,z_4,\,z_1\,z_5,\,z_1\,z_9,\,z_2\,z_3,\,z_2\,z_5,\,z_4\,z_8,\,z_5\,z_8,\\ &z_3\,z_4,\, z_4\,z_5,\,
z_6\,z_7\,z_8,\,z_2\,z_6\,z_7\,z_9,\,z_3\,z_6\,z_7\,z_9\}
\end{aligned}
\end{equation}
The CY hypersurface equation at the symmetric complex structure point becomes:
\begin{equation}
\begin{aligned}
\textmd{equ}_\textmd{rd}=&z_6^3 + z_7^2 + z_1^6\, z_2^{12}\, z_4^{12}\, z_8^6\, z_9^6 + z_2^{12}\, z_4^{18}\, z_5^6\, z_8^6\, z_9^{12} + z_3^{12}\, z_4^6\, z_5^{18}\, z_8^6\, z_9^{12} +\\&
 +\rho_1\, z_1^6\, z_3^{12}\, z_5^{12}\, z_8^6\, z_9^6
+ \rho_2\, z_1^{12}\, z_2^6\, z_3^6\, z_8^6
 + \rho_3\, z_2^6\, z_3^6\, z_4^{12}\, z_5^{12}\, z_8^6\, z_9^{12} +\\&
 + \rho_4\, z_1^6\, z_2^6\, z_3^6\, z_4^6\, z_5^6\, z_8^6\, z_9^6
 + \rho_5\, z_1^4\, z_2^4\, z_3^4\, z_4^4\, z_5^4\, z_6\, z_8^4\, z_9^4
\end{aligned}
\end{equation}

The involution, which is a $\mathbb Z_2$-symmetry of $\Delta_4^{\tilde{X}}$, is given by:
\begin{equation}\label{eq:invo_example3}
 \sigma:\quad (z_2\,,\,z_4)\leftrightarrow(z_3\,,\,z_5)\,.
\end{equation}
Under \eqref{eq:invo_example3} the two dP$_8$ divisors $D_4$ and $D_5$ are mapped onto each other.
Furthermore, as a fixed point set we obtain two $O7$-planes:
\begin{equation}
 O7_1:\quad z_2\,z_4-z_3\,z_5=0\,,\qquad O7_2:\quad z_2\,z_4+z_3\,z_5=0
\end{equation}
with $\chi(O7_1)=\chi(O7_2)=25$ but no $O3$-planes.

\subsection{Example 4}
\label{sec:example4}

In this section we consider a three moduli example of mirror pairs of
CY three-folds $(X_3,\tilde{X}_3)$. The polytope
$\Delta_4^{X}$ and its dual $\Delta_4^{\tilde{X}_3}$ read
\beq
	\Delta_4^{X} =
	\left(
	\begin{array}{rrrr||c}
		6& 0&-1& 2& D_1\\
		6&-1& 0& 2& D_2\\
		3& 1&-1& 1& D_3\\
		3&-1& 1& 1& D_4\\
		0& 0& 1& 0& D_5\\
	    0& 1& 0& 0& D_6\\
        0& 0& 0& 1& D_7\\
      -1& 0& 0&-1& D_8\\
       3& 0& 0& 1 & D_9
	\end{array}
	\right),\qquad
	\Delta_4^{\tilde{X}} =
	\left(
	\begin{array}{rrrr||c}
	-1&-1&-1& 2& \tilde{D}_1\\
	 0&-1&-1&-1& \tilde{D}_2\\
     2&-1&-1&-1&\tilde{D}_3\\
	 2&-1& 5&-1& \tilde{D}_4\\
	2& 5&-1&-1& \tilde{D}_5\\
	2& 5&11&-1& \tilde{D}_6\\
	2&11& 5&-1 & \tilde{D}_7\\
	2&11&11&-1& \tilde{D}_8
\end{array}
	\right)\,.
	\label{eq:delta4ex4}
\eeq
Here we have introduced the toric divisors $D_i$ respectively
$\tilde{D}_i$ corresponding to the points in $\Delta_{4}^{X}$
respectively $\Delta_{4}^{\tilde{X}}$ in the last column. We note
that the last point in $\Delta_{4}^{X}$ is an interior point
on a codimension two face, that has resolves a singularity of the
toric ambient space which also lies in the generic CY
hypersurface $X_3$.
We note that the polytopes $\Delta_4^{X}$ and $\Delta_4^{\tilde{X}}$ are
congruent. Thus the corresponding toric varieties $\mathds{P}_{\Delta}$ and
$\mathds{P}_{\tilde{\Delta}}$ differ only by the action of an orbifold group and
as before
the mirror construction agrees with the GP orbifold
construction of the mirror CY three-fold.

We recall again that it is the CY three-fold $X_3$ of
which we need to calculate the periods on its complex structure moduli
space. Its Hodge numbers and Euler number read
\beq
	h^{(1,1)}=5\,,\quad h^{(2,1)}=185\,,\quad \chi(X_3)=-360\,.
\eeq
After demanding invariance under the GP orbifold group,
only five of the $185$ complex structure moduli remain. The periods
left invariant under this orbifold precisely agree with the periods
of the mirror $\tilde{X}_3$, that we calculate in the next.

All the following topological calculations are performed on the K\"ahler
moduli space of $X_3$. First we note that there are 18
star-triangulation of $\Delta_4^{X}$ yielding 18
CY phase on $X_3$. We chose one phase with  Mori cone
generated by
\bea\label{eq:ellsZ3ex4}
	&\ell^{(1)}=(-1,1,1,0,0,0,2,3,0)\,,\quad \ell^{(2)}=
	(0,0,0,1,-1,1,0,0,-1)\,,&\nn\\
	&\ell^{(3)}=(0,0,1,0,1,-1,0,0,-1)\,,\quad \ell^{(4)}=(0,1,0,-1,1,0,0,0,-1)\,,&\nn\\
	&\ell^{(5)}=(1,-1,-1,0,0,0,0,0,1)\,.&
\eea
Next, we readily calculate by duality between curves and divisor,
the generators $K_i$ of the K\"ahler cone. In the chosen
triangulation the K\"ahler cone is spanned by the generators
\beq \label{eq:Kaehlerconeex4}
\begin{aligned}
K_1=&2\,D_1+2\,D_2+D_3+D_4+D_9\,,\quad
K_2=D_1+D_2+ D_4\,,\quad
K_3= D_1+ D_3\,,\quad\\&\qquad\qquad
K_4= D_1+ D_2\,,\quad
K_5=3\,D_1+2\,D_2+D_3+ D_4+ D_9
\end{aligned}
\eeq
in terms of the toric divisors $D_i$ introduced in \eqref{eq:delta4ex4}.
In terms of this basis we calculate the classical triple intersections as
\bea \label{eq:C0ex4}
	\mathcal{C}_0&=&J_1^3+\frac{3}{2} J_1^2 J_2+\frac{1}{2} J_1 J_2^2+J_1^2 J_3+J_1 J_2 J_3+J_1^2 J_4+J_1 J_2 J_4+J_1 J_3 J_4+\frac{7}{2} J_1^2 J_5+3 J_1 J_2 J_5\nn\\
	&+&\frac{1}{2} J_2^2 J_5+2 J_1 J_3 J_5+J_2 J_3 J_5+2 J_1 J_4 J_5+J_2 J_4 J_5+J_3 J_4 J_5+\frac{7}{2} J_1 J_5^2+\frac{3}{2} J_2 J_5^2+J_3 J_5^2\nn\\
	&+&J_4 J_5^2+\frac{7 J_5^3}{6}\,.
\eea
Analogously we calculate the
other classical intersections \eqref{eq:classic_terms_threefold2} to be
\bea \label{eq:subleadingtermsex4}
	\cK_{11}= -3\,,\,\, \cK_{12}=\cK_{52}= -\frac{1}{2}\,,\,\, \cK_{21}=\cK_{25}= -\frac{3}{2}\,,\,\, \cK_{31}=\cK_{35}=\cK_{41}=\cK_{45}= -1\,,\nn\\
	\cK_{51}=\cK_{55}=\cK_{15} =-\frac{7}{2}\,,\quad
  \cK_j J_j = 3J_1+\frac{3}{2}J_2+J_3+J_4+\frac{41}{12}J_5\ , \qquad
  \cK_0= -\frac{\zeta(3)}{(2 \pi i)^3}360\,\nn\\
\eea
with all other $\cK_{ij}=0$. We note that these relations hold up to integers
as before.

Finally, the intersections \eqref{eq:C0ex4} and
\eqref{eq:subleadingtermsex4} allow us to obtain the prepotential
\eqref{eq:pre_largeV}. The cubic terms are obtained by replacing $J_i\rightarrow -t_i$ in \eqref{eq:C0ex4} and
we obtain
\bea
	F&=&C_0\vert_{J_i=-t_i}
	+\frac{3}{2} t_1^2
	+t_1t_2
	+\frac{1}{2}t_1t_3
	+\frac{1}{2}t_1t_4
	+\frac{7}{2}t_1t_5
	+t_2t_5
	+\frac{1}{2}t_3t_5
	+\frac{7}{4}t_5^2\nn\\
	&&+3t_1+\frac{3}{2}t_2+t_3+t_4+\frac{41}{12}t_5-i\zeta(3)\frac{45}{2\pi^3}
 +  \sum_{\beta} n_\beta^0\,\text{Li}_3(q^\beta)\,,
\eea
where $\beta=(d_1,d_2,d_3,d_4)$ in the basis $\beta^i$ of effective curves
of $H_2(X_3,\mathbf{Z})$ corresponding to the generators
\eqref{eq:ellsZ3ex4} of the Mori cone and $q^\beta=e^{2\pi id_it^i}$ as
before. We also introduced the flat coordinates $t_i$ on the K\"ahler
moduli space of $X_3$ corresponding to the $\beta^i$. The
instanton corrections for the K\"ahler moduli space are determined by
solving the PF equations for $\tilde{X}_3$ along the lines of section
\ref{sec:PFOtoricHS} for the charge vectors \eqref{eq:ellsZ3}.

\subsubsection{The orientifold involution}

The weight matrix for $\Delta_4^{\tilde{X}}$, which was given in \eqref{eq:delta4ex4}, reads as follows:
\begin{equation}
\begin{array}{|c|c|c|c|c|c|c|c|c||c|}
\hline z_1 & z_2 & z_3 & z_4 & z_5 & z_6 & z_7 & z_8 & z_9 & D_{eq_X} \tabularnewline \hline \hline
1 &  0 &  0 &  0 &  1 &  0 &  4 &  6 &  0 &  12\tabularnewline\hline
0 &  1 &  1 &  0 &  1 &  0 &  6 &  9 &  0 &  18\tabularnewline\hline
0 &  1 &  0 &  0 &  0 &  1 &  4 &  6 &  0  & 12\tabularnewline\hline
0 &  0 &  1 &  1 &  0 &  0 &  4 &  6 &  0 &  12\tabularnewline\hline
0 &  0 &  0 &  0 &  0 &  0 &  2 &  3 &  1 & 6\tabularnewline\hline
\end{array}
\label{eq:ex4:weightm}\, .
\end{equation}
The Stanley-Reisner ideal for the phase of the ambient space, which gives rise to two dP$_8$ divisors and one dP$_1$ divisor on the CY and respects the K3-fibration structure of the polytope, is given by:
\begin{equation}
\begin{aligned}
\textmd{SR-ideal}:\quad\{&
z_1\,z_5,\, z_1\,z_6,\, z_2\,z_3,\, z_2\,z_5,\, z_2\,z_6,\, z_2\,z_9,\, z_3\,z_4,\\
& z_3\,z_5,\, z_5\,z_9,
z_7\,z_8\,z_9,\, z_1\,z_4\,z_7\,z_8,\, z_4\,z_6\,z_7\,z_8\}
\end{aligned}
\end{equation}
The CY hypersurface equation at the symmetric complex structure point reads as follows:
\begin{equation}
\begin{aligned}
\textmd{equ}_\textmd{rd}=&
z_7^3+z_8^2
+z_2^6\, z_4^{12}\, z_5^{12}\, z_6^6\, z_9^6
+z_1^6\, z_3^{12}\, z_5^6\, z_6^{12}\, z_9^6
+z_3^6\, z_4^6\, z_5^{12}\, z_6^{12}\, z_9^6+\\&
+\rho_1\, z_1^{12}\, z_2^6\, z_3^{12}\, z_6^6\, z_9^6
+\rho_2\, z_1^6\, z_2^{12}\, z_4^{12}\, z_5^6\, z_9^6
+\rho_3\, z_1^{12}\, z_2^{12}\, z_3^6\, z_4^6\, z_9^6+\\&
+\rho_4\, z_1^6\, z_2^6\, z_3^6\, z_4^6\, z_5^6\, z_6^6\, z_9^6
+\, z_1^4\, z_2^4\, z_3^4\, z_4^4\, z_5^4\, z_6^4\, z_7\, z_9^4
\end{aligned}
\end{equation}

The involution which leaves the ambient variety invariant and maps the two dP$_8$'s at $D_2$ and $D_5$ onto each other is given by:
\begin{equation}
 \sigma:\quad (z_1\,,\,z_2)\leftrightarrow(z_6\,,\,z_5)\,.
\end{equation}
The fixed point set of this involution is given by one $O7$-plane:
\begin{equation}
 O7:\quad z_1\,z_2-z_6\,z_5=0\,
\end{equation}
with $\chi(O7)=24$ and $2\times 1$ plus $2\times 3$ $O3$-planes at
\begin{equation}
\begin{aligned}
&\{z_4=z_1\, z_2+z_5\, z_6=z_9=0\},\quad
\{z_3=z_1\, z_2+z_5\, z_6=z_9=0\},\\&
\{z_4=z_1\, z_2+z_5\, z_6=z_8=0\},\quad
\{z_3=z_1\, z_2+z_5\, z_6=z_8=0\}. 
\end{aligned}
\end{equation}

\begin{footnotesize}
\bibliographystyle{utphys}
\bibliography{complexbib}
\end{footnotesize}
\end{document}